\documentclass[onecolumn,showpacs,preprintnumbers,amsmath,amssymb]{revtex4}

\usepackage{graphicx}
\usepackage{dcolumn}
\usepackage{bm}

\newcommand{\Real}{\mathop{\textrm{Re}}}
\renewcommand{\theequation}{\arabic{section}.\arabic{equation}}

\begin{document}

\preprint{APS/123-QED}

\title{Magnetizability of the relativistic hydrogenlike atom in an arbitrary discrete energy eigenstate: 
Application of the Sturmian expansion of the generalized 
Dirac--Coulomb Green function}

\author{Patrycja Stefa{\'n}ska}
 \email{pstefanska@mif.pg.gda.pl}
\affiliation{Atomic Physics Division, \mbox{Department of Atomic, Molecular and Optical Physics,} Faculty of Applied Physics and Mathematics, Gda{\'n}sk University of Technology,  Narutowicza 11/12, 80--233 Gda{\'n}sk, Poland}


\begin{abstract}
\begin{center}
\textbf{Published as: Phys.\ Rev.\ A 92 (2015) 032504/1-12}
\\*[1ex]
\textbf{doi: 10.1103/PhysRevA.92.032504}\\*[5ex]
\textbf{Abstract} \\*[0.5ex]
\end{center}

The Sturmian expansion of the generalized Dirac--Coulomb Green function
[R.\/~Szmytkowski, J.\ Phys.\ B \textbf{30}, 825 (1997); \textbf{30}, 2747(E) (1997)] is exploited to derive a closed-form expression for the magnetizability of the relativistic one-electron atom in an arbitrary discrete state, with a point-like, spinless and motionless nucleus of charge $Ze$. The result has the form of a double finite sum involving the generalized hypergeometric functions ${}_3F_2$ of the unit argument. Our general expression agrees with formulas obtained analytically earlier by other authors for some particular states of the atom. We present also  numerical values of the magnetizability for some excited states of selected hydrogenlike ions with $1 \leqslant Z \leqslant 137$ and compare them with data available in the literature.
\end{abstract}

\maketitle

\section{\label{sec:I}Introduction}
\setcounter{equation}{0}

The interaction of atoms and molecules with electromagnetic field is arguably one of the most commonly reported physical issues. Not infrequently, a quantitative description of these processes in macroscopic scale comes down to determine the electric and/or magnetic susceptibilities of considered particles. Theoretical investigations of the physical quantities for the simplest systems, like one- and two-electron atoms, were carried out already in the early years of quantum mechanics, and the results were exhaustively presented by Van Vleck \cite{Vleck32}, which discussed the magnetizability of \emph{nonrelativistic\/} hydrogenlike ions. The first \emph{relativistic\/} calculations of this susceptibility for one-electron atoms were conducted in 1974 by Granovsky and Nechet \cite{Gran74} and, independently, by Manakov, Rapoport, and Zapryagaev \cite{Mana74}. In somewhat different ways, they derived an analytical expression for the magnetizability ($\chi$) of the ground state of the Dirac one-electron atom, which contains the hypergeometric function ${}_3F_2$ of the unit argument. Both calculations were based on the perturbation theory combined with the Green function technique. The differences in the two approaches concerned the type and form of the utilized Green function. The first group used the integral representation of the radial parts of the first-order Dirac--Coulomb Green function, whereas the second one exploited the Sturmian expansion of the second-order Dirac--Coulomb Green function, proposed by them at the beginning of the 1970's \cite{Zon72, Mana73}.   A few years later, Manakov \emph{et al.\/} extended their considerations to states with zero radial quantum number \cite{Mana76,Zapr81,Zapr85} (see also Ref.\ \cite{Labz93}). The formula for magnetizability of the atomic ground state was later rederived by Szmytkowski \cite{Szmy02b}, who used the expansion of the first-order generalized Dirac--Coulomb Green function in the Sturmian basis \cite{Szmy97} (for other applications of that powerful technique, see Refs.\ \cite{Szmy04, Szmy02a, Miel06, Szmy11, Stef12, Szmy12, Szmy14}). 

All the aforementioned theoretical investigations of this susceptibility were performed only for some particular states of the atom. To the best of our knowledge, more general analytical calculations of $\chi$, have never been carried out. This prompted us to find the closed-form expression for the magnetizability of the relativistic one-electron atom being in an \emph{arbitrary} discrete energy eigenstate. For this purpose, we exploited the above-mentioned Sturmian expansion of the first-order Dirac--Coulomb Green function, proposed by Szmytkowski in Ref.\ \cite{Szmy97}. 
 
\section{Preliminaries}
\label{II}
\setcounter{equation}{0}
We consider a relativistic hydrogenlike atom placed in a weak, static, uniform magnetic field of induction $\boldsymbol{B}$ (directed along the $z$ axis of a Cartesian coordinate system). With regard to the atom, we assume that the mass and charge of the electron are $m_e$ and $-e$, respectively, and that its nucleus is motionless, point-like, and spinless, with the electronic charge $+Ze$. The energy eigenvalue problem for bound states of such system is established by the Dirac equation
\begin{equation}
\left[
-\mathrm{i}c\hbar\boldsymbol{\alpha}\cdot\boldsymbol{\nabla}
+e c \boldsymbol{\alpha}\cdot\boldsymbol{A}(\boldsymbol{r})+\beta
m_ec^{2}-\frac{Ze^{2}}{(4\pi\epsilon_{0})r}-E
\right]
\Psi(\boldsymbol{r})=0
\label{2.1}
\end{equation}
supplemented by the boundary conditions
\begin{equation}
r \Psi(\boldsymbol{r}) \stackrel{r \to 0}{\longrightarrow}0, 
\qquad \qquad 
r^{3/2} \Psi(\boldsymbol{r}) \stackrel{r \to \infty}{\longrightarrow}0. 
\label{2.2}
\end{equation}
In Eq.\ (\ref{2.1}), $\boldsymbol{\alpha}$ and $\beta$ are the standard Dirac matrices \cite{Schi55}, whereas the vector potential $\boldsymbol{A}(\boldsymbol{r})$ written in a symmetric gauge has the form
\begin{equation}
\boldsymbol{A}(\boldsymbol{r})=\frac{1}{2} \boldsymbol{B}\times\boldsymbol{r}.
\label{2.3}
\end{equation}

From now on, we shall assume that the magnetic field is weak enough to allow us to treat the electron-field interaction operator
\begin{equation}
\hat{H}^{(1)}=e c \boldsymbol{\alpha}\cdot\boldsymbol{A}(\boldsymbol{r})
\label{2.4}
\end{equation}
as a small perturbation of the Dirac--Coulomb Hamiltonian describing an isolated atom. The corresponding zeroth-order bound-state eigenproblem is given by the equation
\begin{equation}
\left[
-\mathrm{i}c\hbar\boldsymbol{\alpha}\cdot\boldsymbol{\nabla}
+\beta m_e c^{2}-\frac{Ze^{2}}{(4\pi\epsilon_{0})r}-E^{(0)}
\right]
\Psi^{(0)}(\boldsymbol{r})=0
\label{2.5}
\end{equation}
and the boundary conditions
\begin{equation}
r \Psi^{(0)}(\boldsymbol{r}) \stackrel{r \to 0}{\longrightarrow}0, 
\qquad \qquad 
r^{3/2} \Psi^{(0)}(\boldsymbol{r}) \stackrel{r \to \infty}{\longrightarrow}0. 
\label{2.6}
\end{equation}

In the absence of external perturbations, the atomic state energy levels are
\begin{equation}
E^{(0)} \equiv E_{n \kappa}^{(0)}=m_ec^2\frac{n+\gamma_{\kappa}}{N_{n \kappa}},
\label{2.7}
\end{equation}
with
\begin{equation}
N_{n\kappa}=\sqrt{n^2+2n\gamma_{\kappa}+\kappa^2}
\label{2.8}
\end{equation}
and
\begin{equation}
\gamma_{\kappa}=\sqrt{\kappa^2-(\alpha Z)^2},
\label{2.9}
\end{equation}
where $n$ is the radial quantum number, the Dirac quantum number $\kappa$ is an integer different from zero, while $\alpha$ denotes the Sommerfeld's fine-structure constant and should not be confused with the Dirac matrix $\boldsymbol{\alpha}$. The eigenfunctions of the Dirac--Coulomb problem (\ref{2.5})--(\ref{2.6}) associated with the eigenvalue $E_{n \kappa}^{(0)}$ are given by
\begin{equation}
\Psi^{(0)}(\boldsymbol{r}) \equiv \Psi_{n \kappa \mu}^{(0)}(\boldsymbol{r}) 
=\frac{1}{r} 
\left(
\begin{array} {c}
P_{n\kappa}^{(0)}(r) \Omega_{\kappa\mu}(\boldsymbol{n}_r) \\ 
\textrm{i} Q_{n\kappa}^{(0)}(r) \Omega_{-\kappa\mu}(\boldsymbol{n}_r) 
\end{array} 
\right),
\label{2.10}
\end{equation}
where $\Omega_{\kappa \mu}(\boldsymbol{n}_r)$ (with $\boldsymbol{n}_r=\boldsymbol{r}/r$, $\kappa=\pm1, \pm2, \ldots$ and $\mu=-|\kappa|+\frac{1}{2}, -|\kappa|+\frac{3}{2}, \ldots , |\kappa|-\frac{1}{2}$) are the orthonormal spherical spinors defined as in Ref.\ \cite{Szmy07}, with the space quantization axis chosen along the external magnetic field direction, while the radial functions are normalized to unity in the sense of
\begin{equation}
\int_0^{\infty} \textrm{d}r 
\left\{
[P_{n\kappa}^{(0)}(r)]^2+[Q_{n\kappa}^{(0)}(r)]^2 
\right\}
=1
\label{2.11}
\end{equation}
and are explicitly given by
\begin{eqnarray}
P_{n\kappa}^{(0)}(r)=\sqrt{\frac{Z}{2a_0} \frac{(1+\epsilon_{n \kappa}) (n+2\gamma_{\kappa}) n!}{N_{n\kappa}^2(N_{n\kappa}-\kappa)\Gamma(n+2\gamma_{\kappa})}}  
\left(
\frac{2Zr}{a_{0}N_{n\kappa}}\right)^{\gamma_{\kappa}}\textrm{e}^{-Zr/a_0N_{n\kappa}}
\nonumber \\
\times 
\left[
L_{n-1}^{(2\gamma_{\kappa})}\left(\frac{2Zr}{a_{0}N_{n\kappa}}\right)
+\frac{\kappa-N_{n\kappa}}{n+2\gamma_{\kappa}}L_{n}^{(2\gamma_{\kappa})}\left(\frac{2Zr}{a_{0}N_{n\kappa}}\right)
\right],
\label{2.12}
\end{eqnarray}
\begin{eqnarray}
Q_{n\kappa}^{(0)}(r)=\sqrt{\frac{Z}{2a_0}\frac{ (1-\epsilon_{n \kappa}) (n+2\gamma_{\kappa}) n!}{N_{n\kappa}^2(N_{n\kappa}-\kappa)\Gamma(n+2\gamma_{\kappa})}} 
\left(\frac{2Zr}{a_{0}N_{n\kappa}}\right)^{\gamma_{\kappa}}  \textrm{e}^{-Zr/a_0N_{n\kappa}}
\nonumber \\
\times 
\left[
L_{n-1}^{(2\gamma_{\kappa})} \left(\frac{2Zr}{a_{0}N_{n\kappa}}\right) 
-\frac{\kappa-N_{n\kappa}}{n+2\gamma_{\kappa}}L_{n}^{(2\gamma_{\kappa})}\left(\frac{2Zr}{a_{0}N_{n\kappa}}\right)
\right]. 
\label{2.13}
\end{eqnarray}
Here $L_{n}^{(\beta)}(\rho)$ is the generalized Laguerre polynomial \cite{Magn66}, $a_0$ denotes, as usual, the Bohr radius, and 
\begin{equation}
\epsilon_{n \kappa}=\frac{E_{n\kappa}^{(0)}}{m_ec^2}=\frac{n+\gamma_{\kappa}}{N_{n\kappa}}.
\label{2.14}
\end{equation}

It is easy to show that the functions from Eq.\ (\ref{2.10}) are adjusted to the perturbation $\hat{H}^{(1)}$ given in Eq.\ (\ref{2.4}), i.e., they diagonalize the matrix  of that perturbation. Utilizing this fact, the solutions of eigenproblem (\ref{2.1})--(\ref{2.2}), to the lowest order in the perturbing field, may be approximated as
\begin{equation}
\Psi(\boldsymbol{r}) \simeq \Psi^{(0)}(\boldsymbol{r})+\Psi^{(1)}(\boldsymbol{r})
\label{2.15}
\end{equation}
and
\begin{equation}
E \simeq E^{(0)}+E^{(1)}.
\label{2.16}
\end{equation}
We assume that the corrections $\Psi^{(1)}(\boldsymbol{r})$ and $E^{(1)}$ are small quantities of the first order in $B=|\boldsymbol{B}|$; they solve the inhomogeneous differential equation
\begin{equation}
\left[
-\textrm{i} c \hbar \boldsymbol{\alpha} \cdot \boldsymbol{\nabla} 
+\beta m_ec^2  
-\frac{Z e^2}{(4\pi \epsilon_0)r} 
-E^{(0)} 
\right]
\Psi^{(1)}(\boldsymbol{r})
=- 
\left[
\frac{1}{2} e c \boldsymbol{{B}} \cdot 
\left(
\boldsymbol{r} \times \boldsymbol{\alpha}
\right)
-E^{(1)} 
\right] 
\Psi^{(0)}(\boldsymbol{r}) 
\label{2.17}
\end{equation}
supplemented by the boundary conditions
\begin{equation}
r \Psi^{(1)}(\boldsymbol{r}) \stackrel{r \to 0}{\longrightarrow}0, 
\qquad \qquad
 r^{3/2} \Psi^{(1)}(\boldsymbol{r}) \stackrel{r \to \infty}{\longrightarrow}0. 
\label{2.18}
\end{equation}
Under the assumption that
\begin{equation}
\int_{\mathbb{R}^3} \textrm{d}^3\boldsymbol{r}\: \Psi^{(0)\dagger}(\boldsymbol{r}) \Psi^{(1)}(\boldsymbol{r})=0,
\label{2.19}
\end{equation}
after carrying out some simple transformations from the standard perturbation theory, we obtain
\begin{equation}
E^{(1)} \equiv E_{n\kappa\mu}^{(1)}=\frac{1}{2} ec \boldsymbol{B} \cdot \int_{\mathbb{R}^3} \textrm{d}^3\boldsymbol{r} \: \Psi_{n \kappa \mu}^{(0)\dagger}(\boldsymbol{r}) 
\left(
\boldsymbol{r} \times \boldsymbol{\alpha} 
\right) 
\Psi_{n \kappa \mu}^{(0)}(\boldsymbol{r})
\label{2.20}
\end{equation}
and
\begin{eqnarray}
\Psi^{(1)}(\boldsymbol{r}) \equiv \Psi_{n \kappa \mu}^{(1)}(\boldsymbol{r})= 
-\int_{\mathbb{R}^3} \textrm{d}^3\boldsymbol{r}' \: \bar{G}^{(0)}(\boldsymbol{r},\boldsymbol{r}') 
\left[ 
\frac{1}{2} e c \boldsymbol{B} \cdot 
\left( 
\boldsymbol{r}' \times \boldsymbol{\alpha}
\right)
-E_{n \kappa \mu}^{(1)} 
\right] 
\Psi_{n \kappa \mu}^{(0)}(\boldsymbol{r}'),
\label{2.21}
\end{eqnarray}
where $\bar{G}^{(0)}(\boldsymbol{r},\boldsymbol{r}')$ is the generalized Dirac--Coulomb Green function associated with the energy level (\ref{2.7}) of an isolated atom. Since it is Hermitian in the sense of
\begin{equation}
\bar{G}^{(0)}(\boldsymbol{r},\boldsymbol{r}')=\bar{G}^{(0)\dagger}(\boldsymbol{r}',\boldsymbol{r})
\label{2.22}
\end{equation}
and satisfies the following orthogonality condition: 
\begin{equation}
\int_{\mathbb{R}^3} \textrm{d}^3\boldsymbol{r} \: \Psi_{n \kappa' \mu}^{(0)\dagger}(\boldsymbol{r}) \bar{G}^{(0)}(\boldsymbol{r},\boldsymbol{r}')=0 
\qquad  
\textrm{for} \:  \kappa'= \pm \kappa,
\label{2.23}
\end{equation}
the formula (\ref{2.21}) for the first order correction to the wave function becomes
\begin{equation}
\Psi_{n \kappa \mu}^{(1)}(\boldsymbol{r})=-\frac{1}{2} e c \boldsymbol{B} \cdot \int_{\mathbb{R}^3} \textrm{d}^3\boldsymbol{r'} \:\bar{G}^{(0)}(\boldsymbol{r},\boldsymbol{r}') 
\left( 
\boldsymbol{r}' \times \boldsymbol{\alpha}
\right) 
\Psi_{n \kappa \mu}^{(0)}(\boldsymbol{r}').
\label{2.24}
\end{equation}

In the rest of this paper, for readability, we shall omit all the subscripts on $\Psi_{n\kappa \mu}^{(0)}$ and $\Psi_{n\kappa \mu}^{(1)}$.

\section{Magnetizability}
\label{III}
\setcounter{equation}{0}

The atomic magnetizability $\chi$ is defined through the relationship
\begin{equation}
\chi=\frac{\mu_0}{4\pi}\frac{\boldsymbol{m}^{(1)} \cdot \boldsymbol{B}}{B^2},
\label{3.1}
\end{equation}
where $\mu_0$ is the permeability of vacuum, while 
\begin{equation}
\boldsymbol{m}^{(1)}=\frac{1}{2}\int_{\mathbb{R}^3} \textrm{d}^3\boldsymbol{r} \: \boldsymbol{r} \times \boldsymbol{j}^{(1)}(\boldsymbol{r})
\label{3.2}
\end{equation}
is an induced magnetic dipole moment of the atom, with
\begin{equation}
\boldsymbol{j}^{(1)}(\boldsymbol{r})=-ec \Psi^{(1)\dagger}(\boldsymbol{r})\boldsymbol{\alpha}\Psi^{(0)}(\boldsymbol{r})-ec \Psi^{(0)\dagger}(\boldsymbol{r})\boldsymbol{\alpha}\Psi^{(1)}(\boldsymbol{r})
\label{3.3}
\end{equation}
being the first-order contribution to an induced current density in the perturbed atomic state $\Psi(\boldsymbol{r})$. Plugging Eq.\ (\ref{3.3}) into (\ref{3.2}), after integrating by parts with use of the boundary conditions satisfied by functions $\Psi^{(0)}(\boldsymbol{r})$ and $\Psi^{(1)}(\boldsymbol{r})$ at infinity, we arrive at
\begin{equation}
\boldsymbol{m}^{(1)}=- e c  \Real \int_{\mathbb{R}^3} \textrm{d}^3\boldsymbol{r} \: \Psi^{(0)\dagger}(\boldsymbol{r}) 
\left(\boldsymbol{r} \times \boldsymbol{\alpha} 
\right) 
\Psi^{(1)}(\boldsymbol{r}).
\label{3.4}
\end{equation}
On substituting the last equation into the definition (\ref{3.1}), utilizing also the formula (\ref{2.24}) and Maxwell's identity $\mu_0 \epsilon_0 c^2=1$, keeping in mind the direction of the perturbing field, i.e. $\boldsymbol{B}=B\boldsymbol{n}_z$, we obtain
\begin{eqnarray}
\chi&=&\frac{1}{2}\frac{e^2}{4\pi\epsilon_0}  \int_{\mathbb{R}^3}  \textrm{d}^3 \boldsymbol{r}  \int_{\mathbb{R}^3}  \textrm{d}^3 \boldsymbol{r'}    
\left[
\boldsymbol{n}_z \cdot 
\left(
\boldsymbol{n}_r \times \boldsymbol{\alpha}
\right) 
\Psi^{(0)}(\boldsymbol{r})
\right]^{\dagger}  
r \bar{G}^{(0)}(\boldsymbol{r}, \boldsymbol{r}') r'  
\left[
\boldsymbol{n}_z \cdot(\boldsymbol{n}_r' \times \boldsymbol{\alpha})\Psi^{(0)}(\boldsymbol{r}') 
\right].
\label{3.5}
\end{eqnarray}
The symbol $\Real$ has been omitted, because the above double integral is obviously real.

To determine the integrals on the right-hand side of Eq.\ (\ref{3.5}), we shall exploit the relation (see Eq.\ (3.1.6) in  \cite{Szmy07})
\begin{eqnarray}
\boldsymbol{n}_z\cdot 
\left( 
\boldsymbol{n}_r \times \boldsymbol{\sigma} 
\right) 
\Omega_{\kappa \mu}(\boldsymbol{n}_r)=
\textrm{i}\frac{4\mu \kappa}{4\kappa^2-1}\Omega_{-\kappa \mu}(\boldsymbol{n}_r)
+\textrm{i}\frac{\sqrt{\left(\kappa+\frac{1}{2}\right)^2-\mu^2}}{|2\kappa+1|}\Omega_{\kappa+1, \mu}(\boldsymbol{n}_r)
-\textrm{i}\frac{\sqrt{\left(\kappa-\frac{1}{2}\right)^2-\mu^2}}{|2\kappa-1|}\Omega_{\kappa-1, \mu}(\boldsymbol{n}_r),
\label{3.6} 
\end{eqnarray}
 and the following partial-wave expansion of the generalized Dirac--Coulomb Green function:
{\footnotesize
\begin{eqnarray}
\bar{G}\mbox{}^{(0)}(\boldsymbol{r},\boldsymbol{r}')
= \frac{4\pi\epsilon_{0}}{e^{2}} 
\sum_{\substack{\kappa'=-\infty \\ (\kappa'\neq0)}}^{\infty}
\sum_{\mu'=-|\kappa'|+1/2}^{|\kappa'|-1/2}\frac{1}{rr'} 
\left(
\begin{array}{cc}
\bar{g}\mbox{}^{(0)}_{(++)\kappa'}(r,r')
\Omega_{\kappa' \mu'}(\boldsymbol{n}_{r})
\Omega_{\kappa' \mu'}^{\dag}(\boldsymbol{n}_{r}^{\prime}) &
-\mathrm{i}\bar{g}\mbox{}^{(0)}_{(+-)\kappa'}(r,r')
\Omega_{\kappa' \mu'}(\boldsymbol{n}_{r})
\Omega_{-\kappa' \mu'}^{\dag}(\boldsymbol{n}_{r}^{\prime}) \\
\mathrm{i}\bar{g}\mbox{}^{(0)}_{(-+)\kappa'}(r,r')
\Omega_{-\kappa' \mu'}(\boldsymbol{n}_{r})
\Omega_{\kappa' \mu'}^{\dag}(\boldsymbol{n}_{r}^{\prime}) &
\bar{g}\mbox{}^{(0)}_{(--)\kappa'}(r,r')
\Omega_{-\kappa' \mu'}(\boldsymbol{n}_{r})
\Omega_{-\kappa' \mu'}^{\dag}(\boldsymbol{n}_{r}^{\prime}) 
\end{array} 
\right).
\label{3.7}
\end{eqnarray}}
Carrying out integrations over the angular variables, and taking advantage of the orthogonality relation for the spherical spinors
\begin{equation}
\oint_{4\pi} \textrm{d}^2\boldsymbol{n}_r \: \Omega_{\kappa \mu}^{\dagger}(\boldsymbol{n}_r) \Omega_{\kappa' \mu'}(\boldsymbol{n}_r)=\delta_{\kappa \kappa'}\delta_{\mu \mu'},
\label{3.8}
\end{equation}
the expression (\ref{3.5}) for the magnetizability takes the form 
\begin{eqnarray}
\chi&=&\sum_{\kappa'}\int_0^{\infty}\textrm{d}r \int_0^{\infty} \textrm{d}r'  
\left( 
\begin{array}{cc} 
Q_{n\kappa}^{(0)}(r)& 
P_{n\kappa}^{(0)}(r)
\end{array} 
\right)   
r \bar{\mathsf{G}}_{\kappa'}^{(0)}(r,r')  r' 
\left( 
\begin{array}{c}
Q_{n\kappa}^{(0)}(r')\\
P_{n\kappa}^{(0)}(r')
\end{array} 
\right)  
\nonumber
\\
&& \times 
\left[ 
\frac{8\kappa^2\mu^2}{\left(4\kappa^2-1\right)^2}\delta_{\kappa',\kappa}
+\frac{\left(\kappa+\frac{1}{2}\right)^2-\mu^2}{2\left(2\kappa+1\right)^2}\delta_{\kappa',-\kappa-1} 
+\frac{\left(\kappa-\frac{1}{2}\right)^2-\mu^2}{2\left(2\kappa-1\right)^2} \delta_{\kappa',-\kappa+1} 
\right],
\label{3.9}
\end{eqnarray}
where
\begin{equation}
\bar{\mathsf{G}}\mbox{}^{(0)}_{\kappa'}(r,r')
=\left(
\begin{array}{cc}
\bar{g}\mbox{}^{(0)}_{(++)\kappa'}(r,r') &
\bar{g}\mbox{}^{(0)}_{(+-)\kappa'}(r,r') \\*[1ex]
\bar{g}\mbox{}^{(0)}_{(-+)\kappa'}(r,r') &
\bar{g}\mbox{}^{(0)}_{(--)\kappa'}(r,r')
\end{array}
\right)
\label{3.10}
\end{equation}
is the radial generalized Dirac--Coulomb Green function associated
with the combined total angular momentum and parity quantum number $\kappa'$. 

Let us rewrite Eq.\ (\ref{3.9}) in the following form:
\begin{equation}
\chi=\chi_{\kappa}+\chi_{-\kappa-1}+\chi_{-\kappa+1},
\label{3.11}
\end{equation}
with the components defined by the action of appropriate Kronecker's deltas in the aforementioned formula. Because the Sturmian expansions of the radial generalized Dirac--Coulomb Green function differ in the cases $\kappa'\neq \kappa$ and $\kappa'=\kappa$ \cite{Szmy97}, 
\begin{equation}
\bar{\mathsf{G}}_{\kappa'}^{(0)}(r,r')
=
\sum_{n'=-\infty}^{\infty}{\frac{1}{\mu_{n'\kappa'}^{(0)}-1}}
\left( 
\begin{array}{c}
S_{n'\kappa'}^{(0)}(r) \\
T_{n'\kappa'}^{(0)}(r)
\end{array} 
\right) 
\left(
\begin{array}{cc} 
\mu_{n'\kappa'}^{(0)}S_{n'\kappa'}^{(0)}(r') & 
T_{n'\kappa'}^{(0)}(r')
\end{array}
\right)
\qquad (\kappa' \neq \kappa),
\label{3.12}
\end{equation}
\begin{eqnarray}
\bar{\mathsf{G}}_{\kappa}^{(0)}(r,r')
&=&
\sum_{\substack{n'=-\infty\\(n'\neq n)}}^{\infty}{\frac{1}{\mu_{n' \kappa}^{(0)}-1}}
\left( 
\begin{array}{c}
S_{n'\kappa}^{(0)}(r) \\
T_{n'\kappa}^{(0)}(r)
\end{array} 
\right)
\left( 
\begin{array}{cc} 
\mu_{n'\kappa}^{(0)} S_{n'\kappa}^{(0)}(r') & 
T_{n'\kappa}^{(0)}(r')\end{array}\right)
\nonumber \\
&& +
\left(
\epsilon_{n\kappa}-\frac{1}{2} 
\right)
\left( 
\begin{array}{c}
S_{n\kappa}^{(0)}(r) \\
T_{n\kappa}^{(0)}(r)
\end{array} 
\right)
\left( 
\begin{array}{cc} 
S_{n\kappa}^{(0)}(r') & 
T_{n\kappa}^{(0)}(r') 
\end{array} 
\right) 
\nonumber \\
&& +
\left( 
\begin{array}{c}
I_{n\kappa}^{(0)}(r) \\
K_{n\kappa}^{(0)}(r)
\end{array} 
\right)
\left( 
\begin{array}{cc} 
S_{n\kappa}^{(0)}(r') &  
T_{n\kappa}^{(0)}(r') 
\end{array} \right)
\nonumber \\
&& +
\left( 
\begin{array}{c}
S_{n\kappa}^{(0)}(r) \\
T_{n\kappa}^{(0)}(r)
\end{array} 
\right)
\left( 
\begin{array}{cc} 
J_{n\kappa}^{(0)}(r') & 
K_{n\kappa}^{(0)}(r')
\end{array} 
\right)
\qquad (\kappa'=\kappa),
\label{3.13}
\end{eqnarray}
the first term on the right-hand side of Eq.\ (\ref{3.11}), i.e. $\chi_{\kappa}$, will be derived with the aid of the formula (\ref{3.13}), whereas the two other constituents will be determined collectively, using Eq.\ (\ref{3.12}). In the last two equations
\begin{eqnarray}
S_{n'\kappa'}^{(0)}(r)=\sqrt{\frac{(1+\epsilon_{n \kappa}) N_{n \kappa}(|n'|+2\gamma_{\kappa'})|n'|!}{2 Z N_{n'\kappa'}(N_{n'\kappa'}-\kappa')\Gamma(|n'|+2\gamma_{\kappa'})}}   
\left(\frac{2Zr}{a_0N_{n\kappa}}\right)^{\gamma_{\kappa'}} 
\textrm{e}^{-Zr/a_0N_{n\kappa}} 
\nonumber \\
\times 
\left[
L_{|n'|-1}^{(2\gamma_{\kappa'})}\left(\frac{2Zr}{a_0N_{n\kappa}}\right) 
+\frac{\kappa'-N_{n'\kappa'}}{|n'|+2\gamma_{\kappa'}}L_{|n'|}^{(2\gamma_{\kappa'})}\left(\frac{2Zr}{a_0N_{n\kappa}}\right)
\right]
\label{3.14}
\end{eqnarray}
and
\begin{eqnarray}
T_{n'\kappa'}^{(0)}(r)=\sqrt{\frac{(1-\epsilon_{n \kappa}) N_{n \kappa} (|n'|+2\gamma_{\kappa'}) |n'|!}{2 Z N_{n'\kappa'}(N_{n'\kappa'}-\kappa')\Gamma(|n'|+2\gamma_{\kappa'})}}   
\left(\frac{2Zr}{a_0N_{n\kappa}}\right)^{\gamma_{\kappa'}}  
\textrm{e}^{-Zr/a_0N_{n\kappa}}
\nonumber \\
\times
\left[
L_{|n'|-1}^{(2\gamma_{\kappa'})}\left(\frac{2Zr}{a_0N_{n\kappa}}\right) 
-\frac{\kappa'-N_{n'\kappa'}}{|n'|+2\gamma_{\kappa'}}L_{|n'|}^{(2\gamma_{\kappa'})}\left(\frac{2Zr}{a_0N_{n\kappa}}\right)
\right]
\label{3.15}
\end{eqnarray}
are the radial Dirac--Coulomb Sturmian functions associated with the hydrogenic discrete state energy level, and
\begin{equation}
\mu_{n'\kappa'}^{(0)}=\frac{|n'|+\gamma_{\kappa'}+N_{n'\kappa'}}{n+\gamma_{\kappa}+N_{n \kappa}},
\label{3.16}
\end{equation}
where
\begin{equation}
N_{n'\kappa'}
=\pm\sqrt{\left(|n'|+\gamma_{\kappa'}\right)^2+(\alpha Z)^2}
=\pm \sqrt{|n'|+2|n'|\gamma_{\kappa'}+\kappa'^2}
\label{3.17}
\end{equation}
is a so-called apparent principal quantum number, which takes the positive values if $n'>0$ and negative if $n'<0$; for $n'=0$, in the definition (\ref{3.17}) one chooses the plus sign if $\kappa'<0$ and the minus sign if $\kappa'>0$. Furthermore, the functions $I_{n\kappa}^{(0)}(r)$, $J_{n\kappa}^{(0)}(r)$ and $K_{n\kappa}^{(0)}(r)$, also appearing in Eq.\ (\ref{3.13}), are defined as \cite{Szmy97}
\begin{equation}
I_{n\kappa}^{(0)}(r)
=\epsilon_{n \kappa} 
\left[
-\left(\kappa+\frac{1}{2\epsilon_{n\kappa}}\right)S_{n\kappa}^{(0)}(r)
+r\left(\frac{m_{e}c(1+\epsilon_{n\kappa})}{\hbar}+\frac{\alpha Z}{r}\right)T_{n\kappa}^{(0)}(r)
\right], 
\label{3.18}
\end{equation}
\begin{equation}
J_{n\kappa}^{(0)}(r)
=\epsilon_{n \kappa} 
\left[
-\left(\kappa-\frac{1}{2\epsilon_{n\kappa}}\right)S_{n\kappa}^{(0)}(r)
+r\left(\frac{m_{e}c(1+\epsilon_{n\kappa})}{\hbar}+\frac{\alpha Z}{r} \right)T_{n\kappa}^{(0)}(r)
 \right], 
\label{3.19}
\end{equation}
\begin{equation}
K_{n\kappa}^{(0)}(r)
=\epsilon_{n \kappa} 
\left[
r\left(\frac{m_{e}c(1-\epsilon_{n\kappa})}{\hbar}-\frac{\alpha Z}{r} \right)S_{n\kappa}^{(0)}(r)
+\left(\kappa+\frac{1}{2\epsilon_{n\kappa}}\right)T_{n\kappa}^{(0)}(r) 
\right]. 
\label{3.20}
\end{equation}

According to the procedure described above, $\chi_{\kappa}$ can be written as a sum
\begin{equation}
\chi_{\kappa}=\frac{8\kappa^2\mu^2}{(4\kappa^2-1)^2} 
\left[
R_{\kappa}^{(\infty)}+R_{\kappa}^{(a)}+R_{\kappa}^{(b)}+R_{\kappa}^{(c)} 
\right],
\label{3.21}
\end{equation}
with the components 
\begin{eqnarray}
R_{\kappa}^{(\infty)}=\sum_{\substack{{n'=-\infty}\\(n' \neq n)}}^{\infty}{\frac{1}{\mu_{n' \kappa}^{(0)}-1}} 
\int_0^{\infty} \textrm{d}r \: r  
\left[
Q_{n\kappa}^{(0)}(r) S_{n'\kappa}^{(0)}(r)+ P_{n\kappa}^{(0)}(r) T_{n'\kappa}^{(0)}(r)
\right] 
\nonumber \\
\times
\int_0^{\infty} \textrm{d}r' \: r' 
\left[
\mu_{n' \kappa}^{(0)} Q_{n\kappa}^{(0)}(r') S_{n'\kappa}^{(0)}(r')+ P_{n\kappa}^{(0)}(r') T_{n'\kappa}^{(0)}(r')
\right],
\label{3.22}
\end{eqnarray}
\begin{eqnarray}
R_{\kappa}^{(a)}= 
\left(\epsilon_{n\kappa}-\frac{1}{2} \right) 
\int_0^{\infty} \textrm{d}r \: r 
\left[
Q_{n\kappa}^{(0)}(r) S_{n\kappa}^{(0)}(r)+ P_{n\kappa}^{(0)}(r) T_{n\kappa}^{(0)}(r)
\right] 
\int_0^{\infty} \textrm{d}r' \: r' 
\left[
Q_{n\kappa}^{(0)}(r') S_{n\kappa}^{(0)}(r')+ P_{n\kappa}^{(0)}(r') T_{n\kappa}^{(0)}(r')
\right], 
\label{3.23}
\end{eqnarray}
\begin{eqnarray}
R_{\kappa}^{(b)}=  
\int_0^{\infty} \textrm{d}r \: r 
\left[
Q_{n\kappa}^{(0)}(r) I_{n\kappa}^{(0)}(r)+ P_{n\kappa}^{(0)}(r) K_{n\kappa}^{(0)}(r)
\right]
\int_0^{\infty} \textrm{d}r' \: r' 
\left[
Q_{n\kappa}^{(0)}(r') S_{n\kappa}^{(0)}(r')+ P_{n\kappa}^{(0)}(r') T_{n\kappa}^{(0)}(r')
\right], 
\label{3.24}
\end{eqnarray}
\begin{eqnarray}
R_{\kappa}^{(c)}=
\int_0^{\infty} \textrm{d}r \: r 
\left[
Q_{n\kappa}^{(0)}(r) S_{n\kappa}^{(0)}(r)+ P_{n\kappa}^{(0)}(r) T_{n\kappa}^{(0)}(r)
\right] 
\int_0^{\infty} \textrm{d}r' \: r' 
\left[
Q_{n\kappa}^{(0)}(r') J_{n\kappa}^{(0)}(r')+ P_{n\kappa}^{(0)}(r') K_{n\kappa}^{(0)}(r')
\right]. 
\label{3.25}
\end{eqnarray}
Exploiting Eqs.\ (\ref{3.18})--(\ref{3.20}) and the relations
\begin{subequations}
\begin{equation}
S_{n\kappa}^{(0)}(r)=\frac{\sqrt{a_0}N_{n\kappa}}{Z}P_{n\kappa}^{(0)}(r),
\label{3.26a}
\end{equation}
\begin{equation}
T_{n\kappa}^{(0)}(r)=\frac{\sqrt{a_0}N_{n\kappa}}{Z}Q_{n\kappa}^{(0)}(r),
\label{3.26b}
\end{equation}
\label{3.26}%
\end{subequations}
with some labor, one can show that 
\begin{equation}
R_{\kappa}^{(a)}+R_{\kappa}^{(c)}=0
\label{3.27}
\end{equation}
and 
\begin{equation}
R_{\kappa}^{(b)}=-4 \epsilon_{n \kappa}\frac{a_0 N_{n\kappa}^2}{Z^2}
\left[ 
\int_0^{\infty} \textrm{d}r \: r P_{n\kappa}^{(0)}(r)Q_{n\kappa}^{(0)}(r) 
\right]^2.
\label{3.28}
\end{equation}
Utilizing Eqs.\ (\ref{2.12}) and (\ref{2.13}), and making use of the recurrence formula for the Laguerre polynomials [see Eq.\ (8.971.5) in \cite{Grad94}]
\begin{equation}
L_n^{(\beta)}(\rho)=L_n^{(\beta+1)}(\rho)-L_{n-1}^{(\beta+1)}(\rho)
\label{3.29}
\end{equation}
and the orthogonality relation \cite[Eq.\ (7.414.3)]{Grad94} 
\begin{equation}
\int_0^{\infty} \textrm{d}\rho \: \rho^{\beta}\textrm{e}^{-\rho}L_m^{(\beta)}(\rho)L_n^{(\beta)}(\rho) 
=\frac{\Gamma(n+\beta+1)}{n!}\delta_{m n} 
\quad \quad 
[\Real \beta>-1],
\label{3.30}
\end{equation}
after taking into account Eq.\ (\ref{2.8}), one gets
\begin{equation}
\int_0^{\infty} \textrm{d}r \: r P_{n\kappa}^{(0)}(r)Q_{n\kappa}^{(0)}(r)
=\frac{\alpha a_0}{4N_{n\kappa}}
\left[
2\kappa(n+\gamma_{\kappa})-N_{n\kappa}
\right].
\label{3.31}
\end{equation}
Employing Eqs.\ (\ref{2.14}) and (\ref{3.31}) reduces the formula (\ref{3.28}) to the form
\begin{equation}
R_{\kappa}^{(b)}=-\frac{\alpha^2 a_0^3 }{Z^2} 
\frac{(n+\gamma_{\kappa})
\left[
2\kappa(n+\gamma_{\kappa})-N_{n \kappa} 
\right]^2}
{4 N_{n\kappa}}.
\label{3.32}
\end{equation}
To find the expression for $R_{\kappa}^{(\infty)}$, we put Eqs.\ (\ref{2.12})--(\ref{2.13}) and (\ref{3.14})--(\ref{3.17}) into Eq.\ (\ref{3.22}) and exploit the relations (\ref{3.29})--(\ref{3.30}). This gives
\begin{equation}
R_{\kappa}^{(\infty)}
=\frac{\alpha^2 a_0^3}{Z^2} 
\frac{N_{n \kappa}\Gamma(n+2\gamma_{\kappa}+1)}{32 n! (N_{n \kappa}-\kappa)}
\sum_{\substack{{n'=-\infty}\\(n' \neq n)}}^{\infty}\frac{|n'|! 
\left(A_{n'}\delta_{|n'|,n-1}+B_{n'}\delta_{|n'|,n}+C_{n'}\delta_{|n'|,n+1} 
\right)
}
{N_{n' \kappa} (N_{n' \kappa}-\kappa) \Gamma(|n'|+2\gamma_{\kappa}+1)}, 
\label{3.33}
\end{equation}
where 
\begin{equation}
A_{n'}=2n^2
\left[
(N_{n\kappa}-\kappa)(N_{n'\kappa}-\kappa)-(n-1)(n+2\gamma_{\kappa}-1) 
\right]
\left[
\kappa(N_{n\kappa}-\kappa)(N_{n\kappa}+N_{n'\kappa})-N_{n'\kappa}(2n+2\gamma_{\kappa}-1)
\right], 
\label{3.34}
\end{equation}
\begin{eqnarray}
B_{n'}&=&(N_{n\kappa}-\kappa)^2
\left[
2(n+\gamma_{\kappa})(2\kappa+N_{n\kappa}-N_{n'\kappa})-N_{n\kappa}-N_{n'\kappa}
\right] 
\nonumber \\
&& \times 
\left\{
(N_{n\kappa}+\kappa)(N_{n'\kappa}-N_{n\kappa})-\epsilon_{n\kappa}
\left[
2(n+\gamma_{\kappa}) (2\kappa+N_{n\kappa}-N_{n'\kappa})-N_{n\kappa}-N_{n'\kappa}
\right] 
\right\},
\label{3.35}
\end{eqnarray}
and
\begin{equation}
C_{n'}=(n+2\gamma_{\kappa}+1)^2(N_{n\kappa}-\kappa)^2(N_{n'\kappa}-N_{n\kappa}-2\kappa)
\left[
(N_{n\kappa}+N_{n'\kappa})(N_{n'\kappa}-N_{n\kappa}-2\kappa)+2n+2\gamma_{\kappa}+1
\right]. 
\label{3.36}
\end{equation}
After somewhat tedious calculations, the expression (\ref{3.33}) may be cast into a much simpler form 
\begin{equation}
R_{\kappa}^{(\infty)}=-\frac{\alpha^2 a_0^3}{Z^2}\frac{1}{4N_{n\kappa}} 
\left[
n(n+\gamma_{\kappa})(n+2\gamma_{\kappa})  
\left(
5n^2+10n\gamma_{\kappa}+2 \gamma_{\kappa}^2+\kappa^2
\right) 
-2\kappa^4(n+\gamma_{\kappa})+\kappa N_{n\kappa}^3 
\right].
\label{3.37}
\end{equation}
Inserting Eqs.\ (\ref{3.27}), (\ref{3.32}) and (\ref{3.37}) into Eq.\ (\ref{3.21}), with some labor we find that the constituent $\chi_{\kappa}$ of the magnetizability is
\begin{eqnarray}
\chi_{\kappa}=-\frac{\alpha^2 a_0^3}{Z^2} \frac{2 \kappa^2 \mu^2}{N_{n \kappa} 
\left(4\kappa^2-1\right)^2}
\left[2\kappa^2(n+\gamma_{\kappa})^3 +(n+\gamma_{\kappa})(5n^2+10n\gamma_{\kappa}+2\gamma_{\kappa}^2-2\kappa^2+1)N_{n\kappa}^2\right.
\nonumber \\
\left.
-\kappa(3n^2+6n \gamma_{\kappa}+4\gamma_{\kappa}^2-\kappa^2)N_{n\kappa} 
\right].
\label{3.38}
\end{eqnarray}

We turn now to the derivation of expressions for the two remaining components of $\chi$. As we have already mentioned, according to formulas given in Eqs.\ (\ref{3.9}) and (\ref{3.11}), their sum may be written as
\begin{eqnarray}
\chi_{-\kappa+1}+\chi_{-\kappa-1}=\sum_{\kappa'}  
\left[
\frac{(2\kappa-1)^2-4\mu^2}{8(2\kappa-1)^2} \delta_{\kappa',-\kappa+1}  
+\frac{(2\kappa+1)^2-4\mu^2}{8(2\kappa+1)^2} \delta_{\kappa',-\kappa-1}  
\right] 
R_{\kappa'}, 
\label{3.39}
\end{eqnarray}
where we have defined
\begin{eqnarray}
R_{\kappa'}=\sum_{n'=-\infty}^{\infty}\frac{1}{\mu_{n' \kappa'}^{(0)}-1}   
\int_0^{\infty}\textrm{d}r \: r
\left[
Q_{n\kappa}^{(0)}(r)S_{n'\kappa'}^{(0)}(r)+P_{n\kappa}^{(0)}(r)T_{n'\kappa'}^{(0)}(r) 
\right] 
\nonumber \\
\times 
\int_0^{\infty} \textrm{d}r' \: r'
\left[
\mu_{n' \kappa'}^{(0)} Q_{n\kappa}^{(0)}(r') S_{n'\kappa'}^{(0)}(r')+P_{n\kappa}^{(0)}(r') T_{n'\kappa'}^{(0)}(r') 
\right].
\label{3.40}
\end{eqnarray}
To tackle the first radial integral on the right-hand side of above equation, we shall exploit Eqs.\ (\ref{3.14}), (\ref{3.15}), (\ref{2.12}), and (\ref{2.13}), with the Laguerre polynomials written in the form
\begin{equation}
L_n^{(\beta)}(\rho)=\sum_{k=0}^{n} \frac{(-)^k}{k!} 
\left( 
\begin{array}{c} 
n+\beta \\
n-k 
\end{array} 
\right) 
\rho^k,
\label{3.41}
\end{equation}
and transform the integration variable according to $x=2Zr/a_0 N_{n\kappa}$, yielding
\begin{eqnarray}
\int_0^{\infty} \textrm{d}r \: r
\left[
Q_{n\kappa}^{(0)}(r) S_{n'\kappa'}^{(0)}(r)+P_{n\kappa}^{(0)}(r) T_{n'\kappa'}^{(0)}(r)
\right]
=2\xi\Gamma(n+2\gamma_{\kappa})\sum_{k=0}^{n}\frac{(-)^k}{k!(n-k)!\Gamma(k+2\gamma_{\kappa}+1)}
\nonumber \\
\times
\int_0^{\infty}\textrm{d}x \: x^{\gamma_{\kappa}+\gamma_{\kappa'}+k+1} \textrm{e}^{-x} 
\left[
(n-k)L_{|n'|-1}^{(2\gamma_{\kappa'})}(x)
-\frac{(N_{n\kappa}-\kappa)(N_{n'\kappa'}-\kappa')}{|n'|+2\gamma_{\kappa'}}L_{|n'|}^{(2\gamma_{\kappa'})}(x)
\right],
\label{3.42}
\end{eqnarray}
with
\begin{equation}
\xi=\sqrt{\frac{\alpha^2 a_0^3}{64 Z^2} 
\frac{N_{n\kappa}n!(n+2\gamma_{\kappa})|n'|!(|n'|+2\gamma_{\kappa'})}
{N_{n'\kappa'} (N_{n\kappa}-\kappa)(N_{n'\kappa'}-\kappa')\Gamma(n+2\gamma_{\kappa})\Gamma(|n'|+2\gamma_{\kappa'})}}.
\label{3.43}
\end{equation}
Utilizing the following formula [\cite{Grad94}, Eq.\ (7.414.11)]
 \begin{equation}
\int_0^{\infty} \textrm{d}\rho \: \rho^{\gamma}\textrm{e}^{-\rho}L_n^{(\beta)}(\rho)
=\frac{\Gamma(\gamma+1)\Gamma(n+\beta-\gamma)}{n! \Gamma(\beta-\gamma)}
=(-)^{n}\frac{\Gamma(\gamma+1)\Gamma(\gamma-\beta+1)}{n!\Gamma(\gamma-\beta-n+1)} 
\qquad 
[\Real(\gamma)>-1]
\label{3.44}
\end{equation}
and again the relation (\ref{3.17}), after some further simplifications we obtain
\begin{eqnarray}
\int_0^{\infty} \textrm{d}r \: r 
\left[
Q_{n\kappa}^{(0)}(r)S_{n'\kappa'}^{(0)}(r)+P_{n\kappa}^{(0)}(r) T_{n'\kappa'}^{(0)}(r) 
\right]
=\frac{2 \xi \Gamma(n+2\gamma_{\kappa})}{(|n'|-1)!(N_{n'\kappa'}+\kappa')} 
\sum_{k=0}^{n}\frac{(-)^k}{k!}\frac{\Gamma(\gamma_{\kappa}+\gamma_{\kappa'}+k+2)}{\Gamma(\gamma_{\kappa'}-\gamma_{\kappa}-k-1)} 
\nonumber \\
\times  
\frac{\Gamma(|n'|+\gamma_{\kappa'}-\gamma_{\kappa}-k-2)}{(n-k)!\Gamma(k+2\gamma_{\kappa}+1)} 
\left[
(n-k)(N_{n'\kappa'}+\kappa')+(\kappa-N_{n \kappa})(|n'|+\gamma_{\kappa'}-\gamma_{\kappa}-k-2)
\right].
\label{3.45}
\end{eqnarray}
Proceeding in a similar way, one gets 
\begin{eqnarray}
\int_0^{\infty} \textrm{d}r \: r  
\left[
\mu_{n' \kappa'}^{(0)}Q_{n\kappa}^{(0)}(r) S_{n'\kappa'}^{(0)}(r)+P_{n\kappa}^{(0)}(r) T_{n'\kappa'}^{(0)}(r) 
\right]
=\frac{\xi \Gamma(n+2\gamma_{\kappa})}{(|n'|-1)!(N_{n'\kappa'}+\kappa')}  
\sum_{p=0}^{n}\frac{(-)^p}{p!}\frac{\Gamma(\gamma_{\kappa}+\gamma_{\kappa'}+p+2)}{\Gamma(\gamma_{\kappa'}-\gamma_{\kappa}-p-1)}
\nonumber \\
\times  
\frac{\Gamma(|n'|+\gamma_{\kappa'}-\gamma_{\kappa}-p-2)}{(n-p)!\Gamma(p+2\gamma_{\kappa}+1)}  
\left\{
(\mu_{n' \kappa'}^{(0)}+1) 
\left[
(n-p)(N_{n'\kappa'}+\kappa')+(\kappa-N_{n \kappa})(|n'|+\gamma_{\kappa'}-\gamma_{\kappa}-p-2)
\right]\right.
\nonumber \\
\left.
-(\mu_{n' \kappa'}^{(0)}-1)
\left[
(\kappa-N_{n \kappa})(N_{n'\kappa'}+\kappa')+(n-p)(|n'|+\gamma_{\kappa'}-\gamma_{\kappa}-p-2)
\right]
\right\}. 
\label{3.46}
\end{eqnarray}
Plugging Eqs.\ (\ref{3.16}), (\ref{3.43}), (\ref{3.45}) and (\ref{3.46}) into Eq.\ (\ref{3.40}), after some rearrangements involving, among others, the identity
\begin{equation}
\gamma_{\kappa'}^2-\gamma_{\kappa}^2=\kappa'^2-\kappa^2,
\label{3.47}
\end{equation}
we arrive at
\begin{eqnarray}
R_{\kappa'}&=&\frac{\alpha^2 a_0^3}{Z^2}\frac{n!\Gamma(n+2\gamma_{\kappa}+1)N_{n\kappa}}{32(N_{n\kappa}-\kappa)}\sum_{k=0}^{n}\sum_{p=0}^{n}
\frac{\mathcal{Z}_{\kappa \kappa'}^{n}(k)\mathcal{Z}_{\kappa \kappa'}^{n}(p)}{\Gamma(\gamma_{\kappa'}-\gamma_{\kappa}-k-1)\Gamma(\gamma_{\kappa'}-\gamma_{\kappa}-p-1)}
\nonumber\\
&& \times 
\sum_{n'=-\infty}^{\infty}\frac{\Gamma(|n'|+\gamma_{\kappa'}-\gamma_{\kappa}-k-2)\Gamma(|n'|+\gamma_{\kappa'}-\gamma_{\kappa}-p-2)}{|n'|!\Gamma(|n'|+2\gamma_{\kappa'}+1) (|n'|+\gamma_{\kappa'}-\gamma_{\kappa}-n)} \frac{\kappa'-N_{n' \kappa'}}{N_{n' \kappa'}} 
\nonumber \\
&&\times  
\left[
(n-k)(N_{n'\kappa'}+\kappa')+(\kappa-N_{n \kappa})(|n'|+\gamma_{\kappa'}-\gamma_{\kappa}-k-2)
\right] 
\nonumber \\
&&\times   
\left\{
(|n'|+\gamma_{\kappa'}-\gamma_{\kappa}-n)
\left[
(\kappa-N_{n \kappa})(N_{n'\kappa'}+\kappa') +(n-p)(|n'|+\gamma_{\kappa'}-\gamma_{\kappa}-p-2)
\right]\right.
\nonumber \\
&& \quad 
\left.
-(N_{n'\kappa'}+N_{n \kappa}) 
\left[
(n-p)(N_{n'\kappa'}+\kappa')+(\kappa-N_{n \kappa})(|n'|+\gamma_{\kappa'}-\gamma_{\kappa}-p-2)
\right]
\right\},
\label{3.48}
\end{eqnarray}
where 
\begin{equation}
\mathcal{Z}_{\kappa \kappa'}^{n}(k)=\frac{(-)^{k}}{k!(n-k)!} \frac{\Gamma(\gamma_{\kappa}+\gamma_{\kappa'}+k+2)}{\Gamma(k+2\gamma_{\kappa}+1)}
\label{3.49}
\end{equation}
and analogously for $\mathcal{Z}_{\kappa \kappa'}^{n}(p)$. The above result may be simplified considerably if in  the series $\sum_{n'=-\infty}^{\infty}(\ldots)$ one collects together terms with the same absolute value of the summation index $n'$ (the Sturmian radial quantum number). Proceeding in that way, after much labor, using Eq.\ (\ref{3.17}) and again the identity (\ref{3.47}), one finds that
\begin{eqnarray}
R_{\kappa'}&=&\frac{\alpha^2 a_0^3}{Z^2}\frac{n!\Gamma(n+2\gamma_{\kappa}+1)N_{n\kappa}}{16(N_{n\kappa}-\kappa)}\sum_{k=0}^{n}\sum_{p=0}^{n} 
\frac{\mathcal{Z}_{\kappa \kappa'}^{n}(k)\mathcal{Z}_{\kappa \kappa'}^{n}(p)}{\Gamma(\gamma_{\kappa'}-\gamma_{\kappa}-k-1)\Gamma(\gamma_{\kappa'}-\gamma_{\kappa}-p-1)} {}
\nonumber \\
&& \times 
\sum_{n'=0}^{\infty}\frac{\Gamma(n'+\gamma_{\kappa'}-\gamma_{\kappa}-k-2)\Gamma(n'+\gamma_{\kappa'}-\gamma_{\kappa}-p-2)}{n'!\Gamma(n'+2\gamma_{\kappa'}+1)(n'+\gamma_{\kappa'}-\gamma_{\kappa}-n)}
\nonumber \\
&& \times 
\left\{
(N_{n\kappa}-\kappa)^2(N_{n\kappa}-\kappa')(n'+\gamma_{\kappa'}-\gamma_{\kappa}-k-2)(n'+\gamma_{\kappa'}-\gamma_{\kappa}-p-2)
\right.
\nonumber \\	
&& \quad 
+(n-p)(N_{n\kappa}-\kappa)(n'+\gamma_{\kappa'}-\gamma_{\kappa}-k-2)(n'+\gamma_{\kappa'}-\gamma_{\kappa}-p-2)(n'+\gamma_{\kappa'}-\gamma_{\kappa}-n)
\nonumber	\\
&&\quad
+n'(n'+2\gamma_{\kappa'}) 
\left[
(n-k)(n-p)(2\kappa+\kappa'-N_{n\kappa})+2(2n-k-p)(N_{n\kappa}-\kappa)
\right]
\nonumber \\
&&\quad 
\left.
-n'(n'+2\gamma_{\kappa'})(n-p)(N_{n\kappa}-\kappa)(n'+\gamma_{\kappa'}-\gamma_{\kappa}-n)	
\right\}.
\label{3.50}
\end{eqnarray}
It is possible to achieve  further simplifications. Firstly, notice that the sum of the two series $\sum_{n'=0}^{\infty}(...)$ formed by using the second and the fourth terms between the curly braces, equals zero. Secondly, we may express the remaining two series in terms of the hypergeometric functions ${}_3F_2$ of the unit argument. Since it holds that \cite{Bail35,Slat66}  
\begin{eqnarray}
 {}_3F_2 
\left(
\begin{array}{c} 
a_1, a_2, a_3\\
b_1, b_2
\end{array}
;1 
\right)
=\frac{\Gamma(b_1)\Gamma(b_2)}{\Gamma(a_1) \Gamma(a_2)\Gamma(a_3)} 
\sum_{n=0}^{\infty} \frac{\Gamma(a_1+n)\Gamma(a_2+n)\Gamma(a_3+n)}{n!\Gamma(b_1+n)\Gamma(b_2+n)}
\nonumber \\  
\left[\Real\left(b_1+b_2-a_1-a_2-a_3 \right)>0\right],
\label{3.51} 
\end{eqnarray}
Eq.\ (\ref{3.50}) becomes
\begin{eqnarray}
R_{\kappa'}&=&\frac{\alpha^2 a_0^3}{Z^2}\frac{n!\Gamma(n+2\gamma_{\kappa}+1)N_{n\kappa}}{16(N_{n\kappa}-\kappa)\Gamma(2\gamma_{\kappa'}+1)} 
\sum_{k=0}^{n}\sum_{p=0}^{n}\mathcal{Z}_{\kappa \kappa'}^{n}(k)\mathcal{Z}_{\kappa \kappa'}^{n}(p)
\nonumber \\
&& \times 
\left[
\frac{(N_{n\kappa}-\kappa')(N_{n\kappa}-\kappa)^2}{\gamma_{\kappa'}-\gamma_{\kappa}-n} 
{}_3F_2 
\left(
\begin{array}{c} 
\gamma_{\kappa'}-\gamma_{\kappa}-k-1,\: 
\gamma_{\kappa'}-\gamma_{\kappa}-p-1,\: 
\gamma_{\kappa'}-\gamma_{\kappa}-n\\
\gamma_{\kappa'}-\gamma_{\kappa}-n+1,\:
2\gamma_{\kappa'}+1
\end{array}
;1 
\right)
\right.
\nonumber \\
&& \quad 
+\frac{(n-k)(n-p)(2\kappa+\kappa'-N_{n\kappa})+2(2n-k-p)(N_{n\kappa}-\kappa)}{\gamma_{\kappa'}-\gamma_{\kappa}-n+1}
\nonumber \\
&&\quad 
\left.
\times {}_3F_2 
\left(
\begin{array}{c} 
\gamma_{\kappa'}-\gamma_{\kappa}-k-1,\: 
\gamma_{\kappa'}-\gamma_{\kappa}-p-1,\: 
\gamma_{\kappa'}-\gamma_{\kappa}-n+1 \\
\gamma_{\kappa'}-\gamma_{\kappa}-n+2,\:
2\gamma_{\kappa'}+1
\end{array}
;1 
\right)
\right].
\label{3.52}
\end{eqnarray}
In the next step, we shall eliminate the first ${}_3F_{2}$ function with the help of the recurrence formula
\begin{eqnarray}
{}_3F_2 
\left( 
\begin{array}{c} 
a_1, a_2, a_3-1\\
a_3,b
\end{array}
;1 
\right)
=-\frac{(a_1-a_3)(a_2-a_3)}{a_3(b-a_3)}  
{}_3F_2 
\left(
\begin{array}{c} 
a_1, a_2, a_3\\
a_3+1,b
\end{array}
;1 
\right)
+\frac{\Gamma(b)\Gamma(b-a_1-a_2+1)}{(b-a_3)\Gamma(b-a_1)\Gamma(b-a_2)}
\nonumber \\ 
\left[\Real(b-a_1-a_2)>-1\right]. 
\label{3.53}
\end{eqnarray}
After some simple algebra, we obtain
\begin{eqnarray}
R_{\kappa'}&=&\frac{\alpha^2 a_0^3}{Z^2} \frac{N_{n \kappa}}{16}\frac{N_{n \kappa}-\kappa}{N_{n \kappa}+\kappa'}
\Bigg[
\mathcal{F}_{\kappa}^{n}(4)+ \frac{n!\Gamma(n+2\gamma_{\kappa}+1)}{(N_{n\kappa}-\kappa)^2(\gamma_{\kappa'}-\gamma_{\kappa}-n+1)\Gamma(2\gamma_{\kappa'}+1)}
\nonumber \\
&& 
\times
\sum_{k=0}^n\sum_{p=0}^n \widetilde{\mathcal{Z}}_{\kappa \kappa'}^{n}(k)\widetilde{\mathcal{Z}}_{\kappa \kappa'}^{n}(p) 
{}_3F_2 
\left(
\begin{array}{c} 
\gamma_{\kappa'}-\gamma_{\kappa}-k-1,\: 
\gamma_{\kappa'}-\gamma_{\kappa}-p-1,\: 
\gamma_{\kappa'}-\gamma_{\kappa}-n+1 \\
\gamma_{\kappa'}-\gamma_{\kappa}-n+2,\:
2\gamma_{\kappa'}+1
\end{array}
;1 
\right)
\Bigg],
\label{3.54}
\end{eqnarray}
where 
\begin{equation}
\widetilde{\mathcal{Z}}_{\kappa \kappa'}^{n}(k)=
\left
[2(N_{n \kappa}-\kappa)+(n-k)(\kappa+\kappa')
\right]
\mathcal{Z}_{\kappa \kappa'}^{n}(k)
\label{3.55}
\end{equation}
and similarly for $\widetilde{\mathcal{Z}}_{\kappa \kappa'}^{n}(p)$, whereas the function $\mathcal{F}_{\kappa}^{n}(M)$ is defined by Eq.\ (A.1). Basing on the analysis carried out in the Appendix, we have
\begin{eqnarray}
\mathcal{F}_{\kappa}^{n}(4)&=&(-)^{n+1}
\sum_{k=n-3}^{n}\frac{(-)^k \Gamma(2\gamma_{\kappa}+k+4) \Gamma(k+4)}{k!\Gamma(2\gamma_{\kappa}+k+1)\Gamma(k-n+4)\Gamma(n-k+1)}
\nonumber \\
&=&
2(2n+2\gamma_{\kappa}+1) 
\left[
3(\gamma_{\kappa}^2-1)-5(n+\gamma_{\kappa})(n+\gamma_{\kappa}+1)
\right].
\label{3.56}
\end{eqnarray}
Thus, the expression for $R_{\kappa'}$ can be rewritten as
\begin{eqnarray}
R_{\kappa'}&=&\frac{\alpha^2 a_0^3}{Z^2} \frac{N_{n \kappa}}{16}\frac{N_{n \kappa}-\kappa}{N_{n \kappa}+\kappa'}
\Bigg[ 
2(2n+2\gamma_{\kappa}+1) 
\left[
3(\gamma_{\kappa}^2-1)-5(n+\gamma_{\kappa})(n+\gamma_{\kappa}+1)
\right]
\nonumber \\
&& +\frac{n!\Gamma(n+2\gamma_{\kappa}+1)}{(N_{n\kappa}-\kappa)^2(\gamma_{\kappa'}-\gamma_{\kappa}-n+1)\Gamma(2\gamma_{\kappa'}+1)} 
\sum_{k=0}^n\sum_{p=0}^n\widetilde{\mathcal{Z}}_{\kappa \kappa'}^{n}(k)\widetilde{\mathcal{Z}}_{\kappa \kappa'}^{n}(p)
\nonumber \\
&&\quad
\times {}_3F_2 
\left(
\begin{array}{c} 
\gamma_{\kappa'}-\gamma_{\kappa}-k-1,\: 
\gamma_{\kappa'}-\gamma_{\kappa}-p-1,\: 
\gamma_{\kappa'}-\gamma_{\kappa}-n+1\\
\gamma_{\kappa'}-\gamma_{\kappa}-n+2,\:
2\gamma_{\kappa'}+1
\end{array}
;1 
\right) 
\Bigg].
\label{3.57}
\end{eqnarray}
If Eq.\ (\ref{3.57}) is inserted into Eq.\ (\ref{3.39}), the sum of the two considered components of the magnetizability is
\begin{eqnarray}
\chi_{-\kappa+1}+\chi_{-\kappa-1}&=&\frac{\alpha^2 a_0^3}{Z^2} \frac{N_{n \kappa}}{128(4\kappa^2-1)^2}
\sum_{\kappa'}\frac{\eta_{\kappa\mu}^{(+)}\delta_{\kappa',-\kappa+1}+\eta_{\kappa\mu}^{(-)}\delta_{\kappa',-\kappa-1}}
{N_{n\kappa}+\kappa'} \Bigg[2(2n+2\gamma_{\kappa}+1)(N_{n \kappa}-\kappa)
\nonumber \\
&& \times
\left
[3(\gamma_{\kappa}^2-1)-5(n+\gamma_{\kappa})(n+\gamma_{\kappa}+1)
\right]
+\frac{n!\Gamma(n+2\gamma_{\kappa}+1)}{(N_{n\kappa}-\kappa)(\gamma_{\kappa'}-\gamma_{\kappa}-n+1)\Gamma(2\gamma_{\kappa'}+1)}
\nonumber \\
&&\quad 
\left. 
\times \sum_{k=0}^n\sum_{p=0}^n \widetilde{\mathcal{Z}}_{\kappa \kappa'}^{n}(k)\widetilde{\mathcal{Z}}_{\kappa \kappa'}^{n}(p)
{}_3F_2 
\left(
\begin{array}{c} 
\gamma_{\kappa'}-\gamma_{\kappa}-k-1,\: 
\gamma_{\kappa'}-\gamma_{\kappa}-p-1,\: 
\gamma_{\kappa'}-\gamma_{\kappa}-n+1\\
\gamma_{\kappa'}-\gamma_{\kappa}-n+2,\:
2\gamma_{\kappa'}+1
\end{array}
;1 
\right) 
\right],
\nonumber
\\
\label{3.58}
\end{eqnarray}%
where $\eta_{\kappa\mu}^{(\pm)}=(4\kappa^2-1)^2-4\mu^2(2\kappa\pm1)^2$. Finally, combining Eqs.\ (\ref{3.38}) and (\ref{3.58}), as the formula in Eq.\ (\ref{3.11}) requires, one finds that the total magnetizability $\chi$ of the Dirac one-electron atom in the state characterized by the set of quantum numbers $\{n, \kappa, \mu\}$, is given by
\begin{eqnarray}
\chi \equiv \chi_{n\kappa\mu}&=&\frac{\alpha^2 a_0^3}{Z^2}\frac{1}{128N_{n \kappa}(4\kappa^2-1)^2}
\left\{
\Theta_{n\kappa\mu}^{(\textrm{I})}+\sum_{\kappa'}
\frac{\eta_{\kappa\mu}^{(+)}\delta_{\kappa',-\kappa+1}+\eta_{\kappa\mu}^{(-)} \delta_{\kappa',-\kappa-1}}{N_{n\kappa}+\kappa'}\right.
\nonumber \\
&&\times
\left[
\Theta_{n\kappa}^{(\textrm{II})}+\frac{n!(n^2+2n\gamma_{\kappa}+\kappa^2)\Gamma(n+2\gamma_{\kappa}+1)}{(N_{n\kappa}-\kappa)(\gamma_{\kappa'}-\gamma_{\kappa}-n+1)\Gamma(2\gamma_{\kappa'}+1)}\sum_{k=0}^n\sum_{p=0}^n \widetilde{\mathcal{Z}}_{\kappa \kappa'}^{n}(k)\widetilde{\mathcal{Z}}_{\kappa \kappa'}^{n}(p)
\right.
\nonumber \\
&&\quad 
\left. 
\left.
\times
{}_3F_2 
\left(
\begin{array}{c} 
\gamma_{\kappa'}-\gamma_{\kappa}-k-1,\: 
\gamma_{\kappa'}-\gamma_{\kappa}-p-1,\: 
\gamma_{\kappa'}-\gamma_{\kappa}-n+1 \\
\gamma_{\kappa'}-\gamma_{\kappa}-n+2,\:
2\gamma_{\kappa'}+1
\end{array}
;1 
\right) 
\right]
\right\},
\label{3.59}
\end{eqnarray}
with
\begin{equation}
\Theta_{n\kappa\mu}^{(\textrm{I})}=-256\kappa^2 \mu^2
\left[
2\kappa^2(n+\gamma_{\kappa})^3+(n+\gamma_{\kappa})(5n^2+10n \gamma_{\kappa}+2\gamma_{\kappa}^2-2\kappa^2+1)N_{n\kappa}^2-\kappa(3n^2+6n \gamma_{\kappa}+4\gamma_{\kappa}^2-\kappa^2)N_{n\kappa}
\right]
\label{3.60}
\end{equation}
and
\begin{equation}
\Theta_{n\kappa}^{(\textrm{II})}=2(2n+2\gamma_{\kappa}+1)(\kappa-N_{n \kappa})N_{n\kappa}^2
\left[
5(n+\gamma_{\kappa})(n+\gamma_{\kappa}+1)-3(\gamma_{\kappa}^2-1)
\right].
\label{3.61}
\end{equation}
The above result gives us the magnetizability of the atom in an \emph{arbitrary} state. 

To verify the correctness of the expression given in Eq.\ (\ref{3.59}), first of all, we should compare its form for some particular states of the atom with counterpart analytical formulas available in the literature. Thus, Eq.\ (\ref{3.59}) written for states with zero radial quantum number $n$ has the form
\begin{eqnarray}
\chi_{0\kappa\mu}&=&\frac{\alpha^2 a_0^3}{Z^2}\frac{\kappa^2}{(4\kappa^2-1)^2} 
\left\{
2 \kappa \mu^2(2\gamma_{\kappa}+1)(2\gamma_{\kappa}^2+\gamma_{\kappa}-\kappa^2) 
+\sum_{\kappa'} \frac{\eta_{\kappa \mu}^{(+)}\delta_{\kappa',-\kappa+1}+\eta_{\kappa \mu}^{(-)}\delta_{\kappa',-\kappa-1}}{32(\kappa-\kappa')}
\right.
\nonumber\\ 
&& \times 
\left[
(\gamma_{\kappa}+1)(2\gamma_{\kappa}+1)(2\gamma_{\kappa}+3)
-\frac{2\Gamma^2(\gamma_{\kappa}+\gamma_{\kappa'}+2)}{(\gamma_{\kappa'}-\gamma_{\kappa}+1) \Gamma(2\gamma_{\kappa}+1)\Gamma(2\gamma_{\kappa'}+1)}
\right.
\nonumber \\
&&\quad 
\left. 
\left. 
\times {}_3F_2 
\left(
\begin{array}{c} 
\gamma_{\kappa'}-\gamma_{\kappa}-1,\:
\gamma_{\kappa'}-\gamma_{\kappa}-1,\:
\gamma_{\kappa'}-\gamma_{\kappa}+1\\
\gamma_{\kappa'}-\gamma_{\kappa}+2,\: 
2\gamma_{\kappa'}+1
\end{array}
;1  
\right)
\right]
\right\}.
\label{3.62}
\end{eqnarray}
If the hypergeometric function is transformed with the aid of the formula
\begin{eqnarray}
{}_3F_2 
\left( 
\begin{array}{c} 
a_1, a_2, a_3 \\
a_3+1,b
\end{array}
;1 
\right)
=-\frac{a_3(b-a_3)}{(a_1-a_3)(a_2-a_3)}
{}_3F_2 
\left( 
\begin{array}{c} 
a_1, a_2, a_3-1 \\
a_3,b
\end{array}
;1 
\right)
+\frac{a_3 \Gamma(b)\Gamma(b-a_1-a_2+1)}{(a_1-a_3)(a_2-a_3)\Gamma(b-a_1)\Gamma(b-a_2)}
\nonumber
\\ 
\left[
\Real(b-a_1-a_2)>-1 
\right], \quad
\label{3.63}
\end{eqnarray}
Eq.\ (\ref{3.62}) becomes
\begin{eqnarray}
\chi_{0\kappa\mu}&=&\frac{\alpha^2 a_0^3}{Z^2}\frac{\kappa^2}{(4\kappa^2-1)^2} 
\left\{
2 \kappa \mu^2(2\gamma_{\kappa}+1)(2\gamma_{\kappa}^2+\gamma_{\kappa}-\kappa^2) 
+\sum_{\kappa'}\frac{\eta_{\kappa\mu}^{(+)}\delta_{\kappa',-\kappa+1}+\eta_{\kappa\mu}^{(-)}\delta_{\kappa',-\kappa-1}}{64}
\right.
\nonumber\\
&& 
\left.
\times 
\frac{(\kappa+\kappa')\Gamma^2(\gamma_{\kappa}+\gamma_{\kappa'}+2)}{(\gamma_{\kappa}-\gamma_{\kappa'})\Gamma(2\gamma_{\kappa}+1)\Gamma(2\gamma_{\kappa'}+1)} 
{}_3F_2 
\left(
\begin{array}{c} 
\gamma_{\kappa'}-\gamma_{\kappa}-1,\: 
\gamma_{\kappa'}-\gamma_{\kappa}-1,\: 
\gamma_{\kappa'}-\gamma_{\kappa}\\
\gamma_{\kappa'}-\gamma_{\kappa}+1,\: 
2\gamma_{\kappa'}+1
\end{array}
;1 
\right) 
\right\}.
\label{3.64}
\end{eqnarray}
The expression in Eq.\ (\ref{3.64}) is identical to the formula derived by Manakov \emph{et al.\/} \cite{Mana76} (see also Ref.\ \cite{Labz93}). A very special case of the class of atomic states discussed above is the ground state. Plugging into Eq.\ (\ref{3.64}) the values of quantum numbers, which characterize this particular state, i.e.: $\kappa{=}-1$ and $\mu{=}\pm 1/2$, we arrive at
\begin{eqnarray}
\chi_{0,-1,\pm\frac{1}{2}}\equiv \chi_g=\frac{\alpha^2 a_0^3}{Z^2} 
\left[
-\frac{(\gamma_{1}+1)(4\gamma_{1}^2-1)}{18}-\frac{\Gamma^2(\gamma_1+\gamma_2+2)}{72(\gamma_2-\gamma_1)\Gamma(2\gamma_1+1)\Gamma(2\gamma_2+1)} 
\right.
\nonumber
\\
\left.
\times {}_3F_2 
\left( 
\begin{array}{c} 
\gamma_2-\gamma_1-1,\: 
\gamma_2-\gamma_1-1,\: 
\gamma_2-\gamma_1 \\
\gamma_2-\gamma_1+1,\: 
2\gamma_2+1
\end{array}
;1  
\right) 
\right]
\label{3.65}
\end{eqnarray}
(the subscript $g$ denotes the atomic ground state), in agreement with the results obtained by Granovsky and Nechet \cite{Gran74}, Zapryagaev \emph{et al.\/} \cite{Mana74,Zapr81, Zapr85} and Szmytkowski \cite{Szmy02b}. 

In addition to the analytical calculations, we also performed numerical tests validating our general expression for $\chi$. From the formulas (\ref{3.59})--(\ref{3.61}), we have found numerical values of the magnetizability for some atomic states, which, in general, are not included in the cases discussed above. We decided not to present here all the results we have obtained numerically, but only a few tables with values of $\chi$ for some states of selected hydrogenlike ions; more comprehensive numerical data will be provided elsewhere. 

\begin{table*} [h]  
\caption{\label{tab:1} Relativistic magnetizabilities $\chi$ (in the units of $\alpha^2 a_0^3$) for excited states $2s_{1/2}$, $2p_{1/2}$ and $2p_{3/2}$ of selected hydrogenlike atoms. The numbers in brackets are the powers of 10 by which the entries are to be multiplied. The inverse of the fine structure constant used is $\alpha^{-1}=137.0359895$ \cite{Cohe88}.}
\begin{ruledtabular}
\begin{tabular}{rccccc}
$Z$ & $2s_{1/2}$ ($\mu=\pm1/2$)     & $2p_{1/2}$ ($\mu=\pm 1/2$)      & $2p_{3/2}$ ($\mu=\pm 1/2$)       & $2p_{3/2}$ ($\mu=\pm 3/2$) \\ \hline
1 & $-6.99972264911$\phantom{$[+4]$} & $+6.67616221345[+4]$          & $-6.67706218447[+4]$ & $-5.99986154687$\phantom{$[+4]$} \\
2 & $-1.74972264896$\phantom{$[+4]$} & $+4.17116407371[+3]$          & $-4.17341378393[+3]$ & $-1.49986154812$\phantom{$[+4]$} \\
3 & $-7.77500426498[-1]$             & $+8.23460556052[+2]$           & $-8.24460266274[+2]$ & $-6.66528216866[-1]$ \\
4 & $-4.37222648379[-1]$             & $+2.60338569906[+2]$           & $-2.60900780126[+2]$ & $-3.74861553109[-1]$ \\
5 & $-2.79722647937[-1]$             & $+1.06524398074[+2]$           & $-1.06884108291[+2]$ & $-2.39861556850[-1]$ \\
10 & $-6.97226441433[-2]$            & $+6.60046515782$\phantom{[+4]} & $-6.69017535244$\phantom{$[+4]$} & $-5.98615880319[-2]$\\
20 & $-1.72226268778[-2]$             & $+3.98344275277[-1]$           & $-4.20554378259[-1]$ & $-1.48617128850[-2]$ \\
30 & $-7.50036819058[-3]$             & $+7.41154649290[-2]$           & $-8.38254014740[-2]$ & $-6.52858808922[-3]$ \\
40 & $-4.09752221369[-3]$             & $+2.14926792043[-2]$           & $-2.68273520170[-2]$ & $-3.61221432676[-3]$ \\
50 & $-2.52240247365[-3]$             & $+7.81880207625[-3]$           & $-1.11280780225[-2]$ & $-2.26259256980[-3]$ \\
60 & $-1.66664558883[-3]$             & $+3.22446344293[-3]$           & $-5.43315437140[-3]$ & $-1.52972409878[-3]$ \\
70 & $-1.15044414929[-3]$             & $+1.41815248741[-3]$           & $-2.96272030456[-3]$ & $-1.08810032306[-3]$ \\
80 & $-8.15096732424[-4]$             & $+6.33820556550[-4]$           & $-1.74663931133[-3]$ & $-8.01753833846[-4]$ \\
90 & $-5.84706741463[-4]$             & $+2.72574815243[-4]$           & $-1.08836158171[-3]$ & $-6.05730479561[-4]$ \\
100 & $-4.19162799559[-4]$             & $+1.02305365512[-4]$           & $-7.04060814202[-4]$ & $-4.65821090171[-4]$ \\
110 & $-2.95439754668[-4]$             & $+2.37832330726[-5]$	          & $-4.64588471291[-4]$ & $-3.62619035155[-4]$ \\
120 & $-1.99037538972[-4]$             & $-8.55758789370[-6]$           & $-3.04875653836[-4]$ & $-2.84450932603[-4]$ \\
130 & $-1.18403008041[-4]$             & $-1.63328357974[-5]$           & $-1.84712577912[-4]$ & $-2.23954406240[-4]$ \\
135 & $-7.69963822053[-5]$             & $-1.36390908592[-5]$           & $-1.18753971243[-4]$ & $-1.98755253510[-4]$ \\
136 & $-6.65087093457[-5]$             & $-1.22662850197[-5]$           & $-9.79667715907[-5]$ & $-1.94057537507[-4]$ \\
137 & $-4.90325647932[-5]$             & $-9.39848208226[-6]$           & $-4.90305759039[-5]$ & $-1.89466232216[-4]$\\
\end{tabular}
\end{ruledtabular}
\end{table*}
\begin{table*}[h]
\caption{\label{tab:2} Relativistic magnetizabilities $\chi$ (in the units of $\alpha^2 a_0^3$) for excited states $3s_{1/2}$, $3p_{1/2}$, $3d_{3/2}$ and $3d_{5/2}$ of selected hydrogenlike atoms. The numbers in brackets are the powers of 10 by which the entries are to be multiplied. The inverse of the fine structure constant used is $\alpha^{-1}=137.0359895$ \cite{Cohe88}.}
\begin{ruledtabular}
\begin{tabular}{rccccc}
$Z$ & $3s_{1/2}$ ($\mu=\pm 1/2$)        & $3p_{1/2}$ ($\mu=\pm 1/2$)     & $3d_{3/2}$ ($\mu=\pm 1/2$)  & $3d_{5/2}$  ($\mu=\pm 5/2$) \\ \hline
1 & $-3.44989970462[+1]$              & $+2.25307933172[+5]$          & $+7.30096461051[+5]$ & $-2.69996919033[+1]$ \\
2 & $-8.62399703486$\phantom{$[+4]$}  & $+1.40745434671[+4]$          & $+4.56262084060[+4]$ & $-6.74969190474$\phantom{$[+4]$} \\
3 & $-3.83233034921$\phantom{$[+4]$}  & $+2.77778599319[+3]$          & $+9.01099752373[+3]$ & $-2.99969190712$\phantom{$[+4]$} \\
4 & $-2.15524698929$\phantom{$[+4]$}  & $+8.77858954325[+2]$          & $+2.85043308441[+3]$ & $-1.68719191045$\phantom{$[+4]$} \\
5 & $-1.37899695508$\phantom{$[+4]$}  & $+3.59018311920[+2]$          & $+1.16716730295[+3]$ & $-1.07969191473$\phantom{$[+4]$} \\
10 & $-3.43996669207[-1]$             & $+2.21512885609[+1]$          & $+7.27553466950[+1]$ & $-2.69691950437[-1]$\\
20 & $-8.52455105946[-2]$      & $+1.31319402403$\phantom{$[+4]$} & $+4.49921809437$\phantom{$[+4]$} & $-6.71920933136[-2]$ \\
30 & $-3.73268574821[-2]$             & $+2.36342363373[-1]$           & $+8.73048112449[-1]$ & $-2.96923316766[-2]$ \\
40 & $-2.05531202447[-2]$             & $+6.48393859475[-2]$           & $+2.69364404698[-1]$ & $-1.65676658799[-2]$ \\
50 & $-1.27866588439[-2]$             & $+2.15149514368[-2]$           & $+1.06764469967[-1]$ & $-1.04930964231[-2]$ \\
60 & $-8.56476385189[-3]$             & $+7.54578251687[-3]$           & $+4.94235261932[-2]$ & $-7.19362395583[-3]$ \\
70 & $-6.01543976349[-3]$             & $+2.37855639091[-3]$           & $+2.53903783119[-2]$ & $-5.20445336430[-3]$ \\
80 & $-4.35638880996[-3]$             & $+3.37246863105[-4]$           & $+1.40358576739[-2]$ & $-3.91372337026[-3]$ \\
90 & $-3.21336037558[-3]$             & $-4.63370108285[-4]$           & $+8.18144539511[-3]$ & $-3.02913068894[-3]$ \\
100 & $-2.38836398846[-3]$            & $-7.34737468813[-4]$           & $+4.95697547231[-3]$ & $-2.39672255620[-3]$ \\
110 & $-1.76739895034[-3]$            & $-7.70466406095[-4]$	         & $+3.08818329331[-3]$ & $-1.92915544081[-3]$ \\
120 & $-1.27793390973[-3]$            & $-6.98531104154[-4]$           & $+1.96109591675[-3]$ & $-1.57388285049[-3]$ \\
130 & $-8.59898050602[-4]$            & $-5.72986988021[-4]$           & $+1.25981073967[-3]$ & $-1.29775473841[-3]$ \\
135 & $-6.38385440164[-4]$            & $-4.92703237724[-4]$           & $+1.01184048473[-3]$ & $-1.18226302618[-3]$ \\
136 & $-5.80895728560[-4]$            & $-4.74119624289[-4]$           & $+9.68493047166[-4]$ & $-1.16069190213[-3]$ \\
137 & $-4.83147083491[-4]$            & $-4.55336155403[-4]$           & $+9.27011600219[-4]$ & $-1.13959546353[-3]$\\
\end{tabular}
\end{ruledtabular}
\end{table*}

The calculations were performed with the value of the inverse of the fine structure constant $\alpha^{-1}=137.0359895$ \cite{Cohe88}, which differs slightly from the current value $137.035999139$ recommended by the Committee on Data for Science and Technology (CODATA) \cite{Nist14}, but which allows us to compare our present results with the previous results of other authors. A significant part of the entries presented in Tables \ref{tab:1} and \ref{tab:2} can be compared with results from Ref.\ \cite{Rutk07} obtained by Rutkowski and Poszwa with the use of a completely different computational method. The agreement turns out to be almost perfect: Comparing over 200 pairs of numbers, we found only three discrepancies. Two of them occur for states $3s_{1/2}$ and $3d_{3/2}$ of the ion with the atomic number $Z=10$, in the ninth and eighth significant digits, respectively, whereas the third one appears for the state $3d_{5/2}$ of the atom with $Z=77$, in the last presented decimal place, so it may be treated as a result of a numerical rounding. Furthermore, the values of $\chi$ for the atomic ground state (not presented here) fully coincide also with those obtained by Szmytkowski \cite{Szmy02b}. For this particular state, there is a critical value of the atomic number, $Z_c$, such that for $Z<Z_c$ the magnetizability is negative, and for $Z \geqslant Z_c$ it assumes positive values. We have found $Z_c=130$, in agreement with predictions from \cite{Szmy02b,Rutk07}, which, however, differs from the values $Z_c=110$ given by Manakov \emph{et al.\/} in Ref.\ \cite{Mana74}, and $Z_c=118$ suggested by the same group a few years later \cite{Zapr85}. 

The change of the sign of the magnetizability occurs not only for the atomic ground state. The same effect, but with $Z_c$ such that for $Z<Z_c$ the magnetizability is positive (this means that the induced magnetic dipole moment is parallel to the external field), and for $Z \geqslant Z_c$ the magnetizability is negative (this corresponds to the situation, when the orientation of the induced magnetic moment and the perturbing magnetic field is anti-parallel), we have observed for states $\mathcal{N}p_{1/2}$ ($\mathcal{N}$ denotes the principal quantum number).

\section{Conclusions}
\label{IV}
\setcounter{equation}{0}

In this paper, we derived analytically a closed-form expression for the magnetizability of the relativistic one-electron atom in an arbitrary discrete state. The result has the form of a double finite sum involving the generalized hypergeometric functions ${}_3F_2$ of the unit argument. From this general formula, we obtained expressions for the susceptibility under consideration, for some particular states of the atom. For states with zero radial quantum number, we reconstructed the corresponding analytical formula found by Manakov \emph{et al.\/} \cite{Mana76,Zapr81,Zapr85} some time ago, from which, consequently, we obtained the expression for the magnetizability of the atomic ground state, in agreement with results predicted by several other authors \cite{Gran74,Mana74,Szmy02b}. This article also contains results of numerical calculations of the magnetizability, performed by us for some other atomic states, which nearly perfectly agree with values given by Rutkowski and Poszwa \cite{Rutk07}. 

This work provides further evidence of the usefulness of the Sturmian expansion of the generalized Dirac--Coulomb Green function \cite{Szmy97}. We have shown that utility of this method goes beyond the study of the atomic ground state -- using it, one may also obtain exact analytical expressions for electromagnetic properties of Dirac hydrogenlike ions in any discrete atomic energy eigenstate.   

\begin{acknowledgments}
I am grateful to Professor R.\ Szmytkowski for very stimulating discussions and for commenting on the manuscript.
\end{acknowledgments}

\appendix

\renewcommand{\theequation}{\Alph{section}.\arabic{equation}}

\section{The simplification of an expression appearing in Eq.\ (\ref{3.54})}

Consider the following expression:
\begin{equation}
\mathcal{F}_{\kappa}^{n}(M)= n!\Gamma(n+2\gamma_{\kappa}+1)
\sum_{k=0}^n\sum_{p=0}^n\frac{(-)^{k+p+1}\Gamma(2\gamma_{\kappa}+k+p+M)}{k!p!(n-p)!(n-k)!\Gamma(k+2\gamma_{\kappa}+1)\Gamma(p+2\gamma_{\kappa}+1)},
\label{A.1}
\end{equation}
where $M$ is the positive integer, satisfying the inequality $M \leqslant n+1$. Let us rewrite the above equation in the form
\begin{equation}
\mathcal{F}_{\kappa}^{n}(M)= n!\Gamma(n+2\gamma_{\kappa}+1) 
\sum_{k=0}^n\frac{(-)^{k+1}}{k!\Gamma(k+2\gamma_{\kappa}+1)\Gamma(n-k+1)} 
\sum_{p=0}^{\infty}\frac{(-)^p\Gamma(2\gamma_{\kappa}+k+p+M)}{p!\Gamma(p+2\gamma_{\kappa}+1)\Gamma(n-p+1)},
\label{A.2}
\end{equation}
where we have used the relation $\Gamma(\zeta+1)=\zeta!$ (for $\zeta\in\mathbb{N}$) and the fact, that the components with $p\in[n+1,\infty)$ equal zero. With the use of the following reflection relation for the gamma function:
\begin{equation}
\Gamma(\zeta)\Gamma(1-\zeta)=\frac{\pi}{\sin{\pi \zeta}},
\label{A.3}
\end{equation}
one has
\begin{equation}
\frac{1}{\Gamma(n-p+1)}=\frac{(-)^{p+1}}{\pi}\lim_{\nu \to n} 
\left[
\sin{(\pi \nu) }\Gamma(p-\nu)
\right].
\label{A.4}
\end{equation}
Inserting the above formula into Eq.\ (\ref{A.2}), we obtain
\begin{equation}
\mathcal{F}_{\kappa}^{n}(M)=\frac{n! \Gamma(n+2\gamma_{\kappa}+1)}{\pi} 
\sum_{k=0}^n  \frac{(-)^{k+1}}{k!\Gamma(k+2\gamma_{\kappa}+1) \Gamma(n-k+1)}\lim_{\nu \to n} 
\left[
\sin{(-\pi \nu)} \sum_{p=0}^{\infty} \frac{ \Gamma(2\gamma_{\kappa}+k+p+M) \Gamma(p-\nu)}{p! \Gamma(p+2\gamma_{\kappa}+1)} 
\right].
\label{A.5}
\end{equation}
Now, it is known from the theory of the hypergeometric functions \cite{Bail35,Slat66} that
\begin{equation}
\sum_{n=0}^{\infty}\frac{\Gamma(a_1+n)\Gamma(a_2+n)}{n!\Gamma(b+n)}
=\frac{\Gamma(a_1)\Gamma(a_2)}{\Gamma(b)}
{}_2F_1 
\left(
\begin{array}{c} 
a_1, a_2 \\ 
b 
\end{array}
;1
\right) 
\qquad 
\left[\Real\left(b-a_1-a_2\right)>0 \right]. 
\label{A.6}
\end{equation}
Consequently, Eq.\ (\ref{A.5}) becomes
\begin{eqnarray}
\mathcal{F}_{\kappa}^{n}(M)&=&\frac{n!\Gamma(n+2\gamma_{\kappa}+1)}{\pi}
\sum_{k=0}^n\frac{(-)^{k+1}}{k!\Gamma(k+2\gamma_{\kappa}+1) \Gamma(n-k+1)}
\nonumber \\
 && \times 
\lim_{\nu \to n}  
\left[ 
\sin{(-\pi \nu)}\frac{\Gamma(2\gamma_{\kappa}+k+M) \Gamma(-\nu)}{\Gamma(2\gamma_{\kappa}+1)} 
{}_2F_1 
\left(
\begin{array}{c}
-\nu,\: 2\gamma_{\kappa}+k+M \\ 
2\gamma_{\kappa}+1
\end{array}
;1 
\right)
\right].
\label{A.7}
\end{eqnarray}
Exploiting the Gauss identity [\cite{Grad94}, Eq.\ (9.122.1)]
\begin{equation}
 {}_2F_1 
\left( 
\begin{array}{c} 
a_1, a_2\\
b
\end{array}
;1 
\right)
=\frac{\Gamma(b)\Gamma(b-a_1-a_2)}{\Gamma(b-a_1)\Gamma(b-a_2)} 
\qquad 
\left[
\Real (b-a_1-a_2)>0 
\right]
\label{A.8}
\end{equation}
and again the relation (\ref{A.3}), we arrive at
\begin{equation}
\mathcal{F}_{\kappa}^{n}(M)=n!\Gamma(n+2\gamma_{\kappa}+1) 
\sum_{k=0}^n\frac{(-)^{k+1}\Gamma(2\gamma_{\kappa}+k+M)}{k!\Gamma(k+2\gamma_{\kappa}+1)\Gamma(n-k+1)\Gamma(-k-M+1)}\lim_{\nu \to n}
\left[
\frac{\Gamma(\nu-k-M+1)}{\Gamma(\nu+1)\Gamma(\nu +2\gamma_{\kappa}+1)} 
\right]. 
\label{A.9}
\end{equation}
Applying once again Eq.\ (\ref{A.3}) and performing the limit procedure, we obtain
\begin{equation}
\mathcal{F}_{\kappa}^{n}(M)=(-)^{n+1} 
\sum_{k=0}^n\frac{(-)^k}{k!}\frac{\Gamma(2\gamma_{\kappa}+k+M)\Gamma(k+M)}{\Gamma(2\gamma_{\kappa}+k+1)\Gamma(k-n+M)\Gamma(n-k+1)}.
\label{A.10}
\end{equation}
In this way, we have eliminated one finite sum from Eq.\ (\ref{A.1}). Next, we note that the only nonzero terms in the above sum are those corresponding to $k\in\{n-M+1, n-M+2, \ldots , n\}$. With this observation, we can finally write that
\begin{equation}
\mathcal{F}_{\kappa}^{n}(M)=(-)^{n+1} 
\sum_{k=n-M+1}^{n}\frac{(-)^k}{k!}\frac{\Gamma(2\gamma_{\kappa}+k+M)\Gamma(k+M)}{\Gamma(2\gamma_{\kappa}+k+1)\Gamma(k-n+M)\Gamma(n-k+1)}.
\label{A.11}
\end{equation}


\begin{thebibliography}{99}
\bibitem{Vleck32}
		J.\ H.\ Van Vleck, 
		\emph{The theory of electric and magnetic susceptibilities} 
		(Oxford University, Oxford, 1932).
\bibitem{Gran74}
		Ya.\ I.\ Granovsky and V.\ I.\ Nechet,
		Static effects in hydrogenlike atoms (Relativistic theory),
		Yad.\ Fiz.\ \textbf{19}, 1290 (1974) 
		[Sov.\ J.\ Nucl.\ Phys.\ \textbf{19}, 660 (1974)].
\bibitem{Mana74}
		N.\ L.\ Manakov, L.\ P.\ Rapoport, and S.\ A.\ Zapryagaev,
		Relativistic electromagnetic susceptibilities of hydrogen-like atoms,
		J.\ Phys.\ B \textbf{7}, 1076 (1974).
\bibitem{Zon72}
		B.\ A.\ Zon, N.\ L.\ Manakov, and L.\ P.\ Rapoport,
		Coulomb Green function in the x-representation and the relativistic polarizability of hydrogen atom,
		Yad.\ Fiz. \textbf{15}, 508 (1972)
		[Sov.\ J.\ Nucl.\ Phys.\ \textbf{15}, 282 (1972)].
\bibitem{Mana73}
		N.\ L.\ Manakov, L.\ P.\ Rapoport, and S.\ A.\ Zapryagaev,
		Sturmian expansion of the relativistic Coulomb Green function,
		Phys.\ Lett.\ A \textbf{43}, 139 (1973).
\bibitem{Mana76} 
		N.\ L.\ Manakov and S.\ A.\ Zapryagaev,
		A reduced Green function of the Dirac equation with a Coulomb potential. Second order Zeeman effect,
		Phys.\ Lett.\ A \textbf{58}, 23 (1976).
\bibitem{Zapr81}
		S.\ A.\ Zapryagaev and N.\ L.\ Manakov,
		The use of the Coulomb Green functions to the studies of relativistic and correlation effects in highly stripped ions,
		Izv.\ Akad.\ Nauk SSSR, Ser.\ Fiz.\ \textbf{45}, 2336 (1981) (in Russian).
\bibitem{Zapr85}
		S.\ A.\ Zapryagaev, N.\ L.\ Manakov, and V.\ G.\ Pal'chikov,
		\emph{Theory of multi-charged ions with one and two electrons},
		(Energoatomizdat, Moscow, 1985) (in Russian).
\bibitem{Labz93} 
		L.\ N.\ Labzovsky, G.\ L.\ Klimchitskaya, and Yu.\ Yu.\ Dmitriev,
		\emph{Relativistic effects in the spectra of atomic systems} 
		(Institute of Physics, Bristol, 1993).
\bibitem{Szmy02b}
   R.\ Szmytkowski,
   Magnetizability of the relativistic hydrogen-like atom:
   application of the Sturmian expansion of the first-order
   Dirac--Coulomb Green function,
   J.\ Phys.\ B \textbf{35}, 1379 (2002). 
\bibitem{Szmy97}
   R.\ Szmytkowski,
   The Dirac--Coulomb Sturmians and the series expansion of the
   Dirac--Coulomb Green function: Application to the relativistic
   polarizability of the hydrogen-like atom,
   J.\ Phys.\ B \textbf{30}, 825 (1997);
	 \textbf{30}, 2747(E) (1997); 
   Addendum, arXiv:physics/9902050.
\bibitem{Szmy04}
   R.\ Szmytkowski and K.\ Mielewczyk,
   Gordon decomposition of the static dipole polarizability of the
   relativistic hydrogen-like atom: Application of the Sturmian
   expansion of the first-order Dirac--Coulomb Green function,
   J.\ Phys.\ B \textbf{37}, 3961 (2004). 
\bibitem{Szmy02a}
   R.\ Szmytkowski,
   Dynamic polarizability of the relativistic hydrogenlike atom:
   Application of the Sturmian expansion of the Dirac--Coulomb Green
   function,
   Phys.\ Rev.\ A \textbf{65}, 012503 (2001). 
\bibitem{Miel06}
   K.\ Mielewczyk and R.\ Szmytkowski,
   Stark-induced magnetic anapole moment in the ground state of the
   relativistic hydrogenlike atom: Application of the Sturmian 
   expansion of the generalized Dirac--Coulomb Green function,
   Phys.\ Rev.\ A \textbf{73}, 022511 (2006);
	 \textbf{73}, 039908(E) (2006).
\bibitem{Szmy11}
   R.\ Szmytkowski and P.\ Stefa{\'n}ska,
   Comment on `Four-component relativistic theory for NMR parameters:
   Unified formulation and numerical assessment of different
   approaches' [J. Chem. Phys. \textbf{130}, 144102 (2009)],
   arXiv:1102.1811.
\bibitem{Stef12}
   P.\ Stefa{\'n}ska and R.\ Szmytkowski,
   Electric and magnetic dipole shielding constants for the ground
   state of the relativistic hydrogen-like atom: Application of the
   Sturmian expansion of the generalized Dirac--Coulomb Green
   function,
   Int.\ J.\ Quantum Chem.\ \textbf{112}, 1363 (2012). 
\bibitem{Szmy12}
   R.\ Szmytkowski and P.\ Stefa{\'n}ska, 
   Magnetic-field-induced electric quadrupole moment in the ground
   state of the relativistic hydrogenlike atom: Application of the
   Sturmian expansion of the generalized Dirac--Coulomb Green
   function, 
   Phys.\ Rev.\ A \textbf{85}, 042502 (2012). 
\bibitem{Szmy14}
   R.\ Szmytkowski and P.\ Stefa{\'n}ska, 
   Electric-field-induced magnetic quadrupole moment in the ground
   state of the relativistic hydrogenlike atom: Application of the
   Sturmian expansion of the generalized Dirac--Coulomb Green
   function, 
   Phys.\ Rev.\ A \textbf{89}, 012501 (2014). 
\bibitem{Schi55}
	See, for example,
	L.\ I.\ Schiff,
	\emph{Quantum Mechanics}, 2nd ed.
  (McGraw-Hill, New York, 1955). 
\bibitem{Szmy07}
   R.\ Szmytkowski,
   Recurrence and differential relations for spherical spinors,
   J.\ Math.\ Chem.\ \textbf{42}, 397 (2007).
\bibitem{Magn66}
   W.\ Magnus, F.\ Oberhettinger, and R.\ P.\ Soni,
   \emph{Formulas and Theorems for the Special Functions of Mathematical
   Physics}, 3rd ed.
   (Springer, Berlin, 1966).
\bibitem{Grad94}
   I.\ S.\ Gradshteyn and I.\ M.\ Ryzhik,
   \emph{Table of Integrals, Series, and Products}, 5th ed.
   (Academic, San Diego, 1994).
\bibitem{Bail35}
   W.\ N.\ Bailey,
   \emph{Generalized Hypergeometric Series}
   (Cambridge University, Cambridge, England, 1935)
	 [reprint: Hafner, New York, 1972].
\bibitem{Slat66}
   L.\ J.\ Slater,
   \emph{Generalized Hypergeometric Functions}
   (Cambridge University, Cambridge, England, 1966).
\bibitem{Cohe88} 
	E.\ R.\ Cohen and B.\ N.\ Taylor,
	The 1986 CODATA recommended values of the fundamental physical constants,
	J.\ Phys.\ Chem.\ Ref.\ Data \textbf{17}, 1795 (1988).
\bibitem{Nist14} CODATA Internationally recommended 2014 values of the fundamental physical constants,
	\textrm{http://physics.nist.gov/cuu/Constants/index.html}.
\bibitem{Rutk07} 
	A.\ Rutkowski  and A.\ Poszwa,
	Static dipole magnetic susceptibilities of relativistic hydrogenlike
	atoms: A semianalytical approach, 
	Phys.\ Rev.\ A \textbf{75}, 033402 (2007). 
\end{thebibliography}
\end{document}